\documentclass[lettersize,journal]{IEEEtran}
\usepackage{amsmath,amsfonts}
\usepackage{algorithmic}
\usepackage{array}
\usepackage[caption=false,font=footnotesize,labelfont=rm,textfont=rm]{subfig}
\usepackage{textcomp}
\usepackage{stfloats}
\usepackage{url}
\usepackage{verbatim}
\usepackage{graphicx}
\def\BibTeX{{\rm B\kern-.05em{\sc i\kern-.025em b}\kern-.08em
    T\kern-.1667em\lower.7ex\hbox{E}\kern-.125emX}}
\usepackage{balance}

\usepackage{xcolor}
\usepackage{framed}
\usepackage{algorithm}

\begin{document}
\title{AI-empowered Channel Estimation for Block-based Active IRS-enhanced Hybrid-field IoT Network}
\author{Yan Wang, Feng Shu, Xianpeng Wang, Minghao Chen, Riqing Chen, Liang Yang, Junhui Zhao
\thanks{Manuscript created October, 2020; 
This work was supported in part by the National Natural Science Foundation of China under Grant U22A2002, and by the Hainan Province Science and Technology Special Fund under Grant ZDYF2024GXJS292; in part by the Scientific Research Fund Project of Hainan University under Grant KYQD(ZR)-21008; in part by the Collaborative Innovation Center of Information Technology, Hainan University, under Grant XTCX2022XXC07; in part by the National Key Research and Development Program of China under Grant 2023YFF0612900. (Corresponding author: Feng Shu).}
\thanks{Yan Wang, Feng Shu, Xianpeng Wang, and Minghao Chen are with the School of Information and Communication Engineering, Hainan University, Haikou 570228, China (e-mail: yanwang@hainanu.edu.cn; shufeng0101@163.com; wxpeng2016@hainanu.edu.cn; chenminghao2023@126.com).}
\thanks{Feng Shu is also with the School of Electronic and Optical Engineering, Nanjing University of Science and Technology, Nanjing 210094, China, and with the School of Mechanical and Electrical Engineering, Hainan Vocational University of Science and Technology, Haikou 571126, China.}
\thanks{Riqing Chen is with the College of Computer and Information Science, Fujian Agriculture and Forestry University, Fuzhou 350002, China, and also with Fujian Key Lab of Agricultural IoT Applications, Sanming University, Sanming, Fujian, 365004, China (e-mail: riqing.chen@fafu.edu.cn).}
\thanks{Liang Yang is with the College of Computer Science and Electronic Engineering, Hunan University, Changsha 410082, China (e-mail: liangy@hnu.edu.cn).}
\thanks{Junhui Zhao is with the School of Electronic and Information Engineering, Beijing Jiaotong University, Beijing 100044, China (e-mail: junhuizhao@hotmail.com).}
}


\maketitle

\begin{abstract}
In this paper, channel estimation (CE) for uplink hybrid-field communications involving multiple Internet of Things (IoT) devices assisted by an active intelligent reflecting surface (IRS) is investigated.
Firstly, to reduce the complexity of near-field (NF) channel modeling and estimation between IoT devices and active IRS, a sub-blocking strategy for active IRS is proposed. 
Specifically, the entire active IRS is divided into multiple smaller sub-blocks, so that IoT devices are located in the far-field (FF) region of each sub block, while also being located in the NF region of the entire active IRS.
This strategy significantly simplifies the channel model and reduces the parameter estimation dimension by decoupling the high-dimensional NF channel parameter space into low dimensional FF sub channels.
Subsequently, the relationship between channel approximation error and CE error with respect to the number of sub blocks is derived, and the optimal number of sub blocks is solved based on the criterion of minimizing the total error.
In addition, considering that the amplification capability of active IRS requires power consumption, a closed-form expression for the optimal power allocation factor is derived.
To further reduce the pilot overhead, a lightweight CE algorithm based on convolutional autoencoder (CAE) and multi-head attention mechanism, called CAEformer, is designed.
The Cram$\acute{e}$r-Rao lower bound is derived to evaluate the proposed algorithm's performance.
Finally, simulation results demonstrate 
the proposed CAEformer network significantly outperforms the conventional least square and minimum mean square error scheme in terms of estimation accuracy.
\end{abstract}


\section{Introduction}

\IEEEPARstart{A}{gainst} the backdrop of the continuous evolution of wireless communication technologies, intelligent reflecting surfaces (IRSs) \cite{Shu2021Enhanced} have paved a novel avenue for enhancing system performance by virtue of their capability to flexibly manipulate electromagnetic waves. However, the performance gains brought about by IRSs are highly contingent upon channel state information (CSI) \cite{Wang2024Enhanced}. Consequently, the channel estimation (CE) issue within IRS-assisted networks has emerged as a central focus of academic inquiry \cite{Sun2023Pilot}\cite{Zhu2024Channel}.

\subsection{Prior Works}

Currently, numerous algorithms have centered their attention on the CE issue in IRS-aided networks. 
For instance, in \cite{Zhou2022Channel}, Zhou et al. proposed a novel low-pilot-overhead cascaded CE methodology. 
This approach capitalized on the inherent sparsity and inter-channel correlations within multi-user (MU) cascaded mmWave MISO networks to enable high-precision CE.
Moreover, an IRS-aided low-earth orbit (LEO) satellite wireless network was introduced in \cite{Zheng2022Intelligent}. 
By harnessing the controllable phase adjustments of numerous passive reflecting elements, the system achieved adaptive beamforming, thereby mitigating the channel impairments induced by dynamic propagation conditions between high-mobility satellites (SATs) and ground nodes (GNs).
Building on this foundation, the authors in \cite{Zheng2022Intelligent} introduced an effective transmission protocol tailored for CE. 
This protocol operates via decentralized and autonomous processing at the SAT and GN, thereby significantly lowering the system's implementation complexity.
Concurrently, the authors of \cite{You2020Channel} devised an IRS reflection matrix and a passive beamforming protocol tailored for data transmission, addressing the CE challenge. 
This design iteratively estimates the IRS element-wise channel parameters across sequential time blocks via hierarchical grouping and partitioning within the IRS structure.
In contrast to the conventional research approach, which involves estimating channels first and subsequently optimizing system parameters, the authors of \cite{Jiang2021Learning} proposed a machine learning (ML)-based methodology.
This approach directly optimizes the beamformer at the base station (BS) and the phase shift matrix at IRS in accordance with system objectives. 
More precisely, it utilizes a ML network to map the received pilot signals to an optimized system setup.
Additionally, it adopts a graph neural network (GNN) framework to model the interdependencies among diverse users within a cellular network.

Previously, relevant research predominantly focused on CE algorithms within the context of passive IRS-aided communication scenarios \cite{Sun2025Channel}. 
However, to effectively mitigate the multiplicative fading effect that degrades communication performance, active IRS emerged and has rapidly become a focal area of research \cite{Peng2025Active}. 
Consequently, researchers have gradually shifted their attention to the development of CE methods tailored for active IRS-aided wireless systems.
For instance, the authors of \cite{Wang2024Power} put forward a method for separate estimation of the direct channel and cascaded channels in an active IRS-assisted uplink Internet of Things (IoT) network. 
This work not only derived the traditional least squares (LS) estimator but also ingeniously integrated convolutional neural networks (CNNs) into the CE process. 
Simulation outcomes demonstrated that the deep learning-based CE algorithm significantly outperformed classical LS and minimum mean square error (MMSE) algorithms in terms of estimation accuracy.
Following that, the authors of \cite{Lin2021Tensor-Based} systematically investigated the CE issue in a hybrid IRS-assisted MIMO OFDM system. 
Considering the distinctive structure of hybrid IRSs, the researchers broke down the complex cascaded CE problem into two separate subproblems: retrieving the channel from the mobile station (MS) to the IRS and the channel from BS to IRS.
By capitalizing on the sparse nature of high-frequency channel propagation, they represented the training signals as a multidimensional tensor structured according to the canonical polyadic decomposition (CPD) framework, wherein certain fibers or slices of the tensor were intentionally omitted or left incomplete.
Simulation results indicated that the proposed scheme exhibited notable advantages with respect to estimation accuracy, interference resilience, and computational efficiency. 
Notably, the enhancements in its performance were particularly evident under scenarios characterized by a restricted count of active elements.

Constrained by the combined characteristics of low-dimensional antenna array architectures and relatively low operating frequency bands, the effective coverage area of traditional wireless near-field (NF) communication systems is typically confined to distances on the order of a few meters or even centimeters \cite{Feng2021Near-Field}. 
However, to align with the technological evolution and performance requirements of sixth-generation (6G) mobile communication networks, future communication systems will adopt larger-scale antenna aperture designs and operate at higher frequency bands, e.g., millimeter-wave, and terahertz spectrums \cite{Liu2025Near-Field}. 
This shift will significantly amplify the prominence of NF communication characteristics within the system \cite{Pan2021RIS-Aided}.
With the integration of emerging technologies, including IRS, extremely large-scale MIMO systems, dynamically reconfigurable antenna arrays, and cell-free distributed architectures, NF communication scenarios will become increasingly pervasive across all tiers of future wireless networks \cite{Ting2025Adaptive}. 
It is crucial to note, however, that the electromagnetic propagation characteristics in the radiative NF region exhibit notable disparities compared to those in the conventional far-field (FF) \cite{Li2025Reinforcement}. 
The enhanced spatial non-stationarity, pronounced spherical wavefront attenuation, and intensified multipath coupling effects within the NF region pose formidable challenges to the precise estimation of channel parameters.

In summary, the existing body of research predominantly concentrates on the design and optimization of channel estimation algorithms tailored for FF communication scenarios assisted by passive IRSs. 
In contrast, for NF or hybrid-field communication systems based on active IRS architectures, the exploration of leveraging deep learning techniques to achieve intelligent estimation of channel parameters remains in its nascent stages. 
Consequently, the corpus of relevant research findings in this specific domain is relatively scarce.

\subsection{Our Contributions}
To the best of our knowledge, there is relatively limited research on the application of convolutional autoencoder (CAE) \cite{Guo2025Unsupervised} and multi-head attention mechanism (MHAM) for CE in active IRS-assisted MU hybrid-field communication systems, especially considering both active IRS and hybrid-field communication.
To fill in this gap and overcome the challenges faced by large artificial intelligence models (LAMs) in practical wireless communication environments, including computational resource constraints and the need for efficient training methods, a lightweight CE framework is proposed. Our primary contributions are summarized as follows.

\begin{enumerate}
  \item To begin with, a block-based FF channel model is proposed to approximate the NF channel model. In this model, the entire active IRS is partitioned into multiple small sub-blocks. In this way, we could reasonably assume that the user is located in the FF region of each small sub-block while being in the NF region of the entire active IRS. This partitioning decouples the high-dimensional NF parameter space into low-dimensional FF sub channels, significantly simplifying the channel model and reducing the parameter estimation dimension. The relationship between the channel approximation error and the CE error with respect to the number of sub-blocks is mathematically derived. Subsequently, the optimal number of sub-blocks is determined by minimizing the total error.
  \item Unlike passive IRSs, active IRSs require power consumption due to their amplification capability. This necessitates optimizing the total power allocation between IoT devices and the active IRS to minimize the channel estimation error. To achieve this, a closed-form expression for the optimal power allocation factor (PAF) is derived. Furthermore, the Cram$\acute{e}$r-Rao lower bound (CRLB) is derived to assess the performance of the proposed CE algorithm.
  \item To further reduce pilot overhead, a lightweight CE framework termed CAEformer is proposed. The framework integrates a convolutional feature extractor with a compact MHAM to construct a hybrid encoder-decoder architecture. Unlike traditional CAEs, which rely solely on local receptive fields, or pure Transformer models \cite{Guo2024Parallel} that require large-scale sequence embeddings, the proposed CAEformer adopts a shallow structural design. By synergistically combining the efficiency of convolutional operations in local feature extraction with the capability of Transformers in modeling long-range dependencies, the proposed CAEformer achieves collaborative capture of both local and global channel characteristics. Simulation results demonstrate that the proposed CAEformer framework significantly outperforms conventional methods such as LS, MMSE, CNN \cite{Wang2024Power}, and deep residual network (DRN) \cite{Wang2024Enhanced} in CE accuracy, while effectively balancing computational complexity and estimation performance.
\end{enumerate}

\vspace{-8pt}
\subsection{Organization and Notation}

The subsequent sections of this paper are organized as described below.
In Section II, the system model of an active IRS-enhanced hybrid-field IoT network is established. 
A block-based FF channel model is proposed in Section III to approximate the NF channel model.
In Section IV, the CRLB is derived.
Simulation results and conclusions are provided in Section V and Section VI, respectively.

Notations: Throughout the paper, vectors and matrices are denoted by boldface lower case and upper case letters, respectively. Lower case letters are employed to represent scalars, while $(\cdot)^{H}$ stands for conjugate and transpose operation. 
The sign $(\cdot)^{T}$ denotes the transpose operation.
The signs $\|\cdot\|_2$ and $\|\cdot\|_F$ stand for the 2-norm and $F$-norm, respectively. 
The notation $\mathbb{E}\{\cdot\}$ denotes the expectation operation. $\hat{[\;]}$ represents the estimation operation. 
$\lvert \cdot \rvert$ represents the operation of taking absolute values.
The symbol $\otimes$ denotes the Kronecker product.


\section{System model}
  
\begin{figure}[h]
\centering
\includegraphics[width=3.1in]
{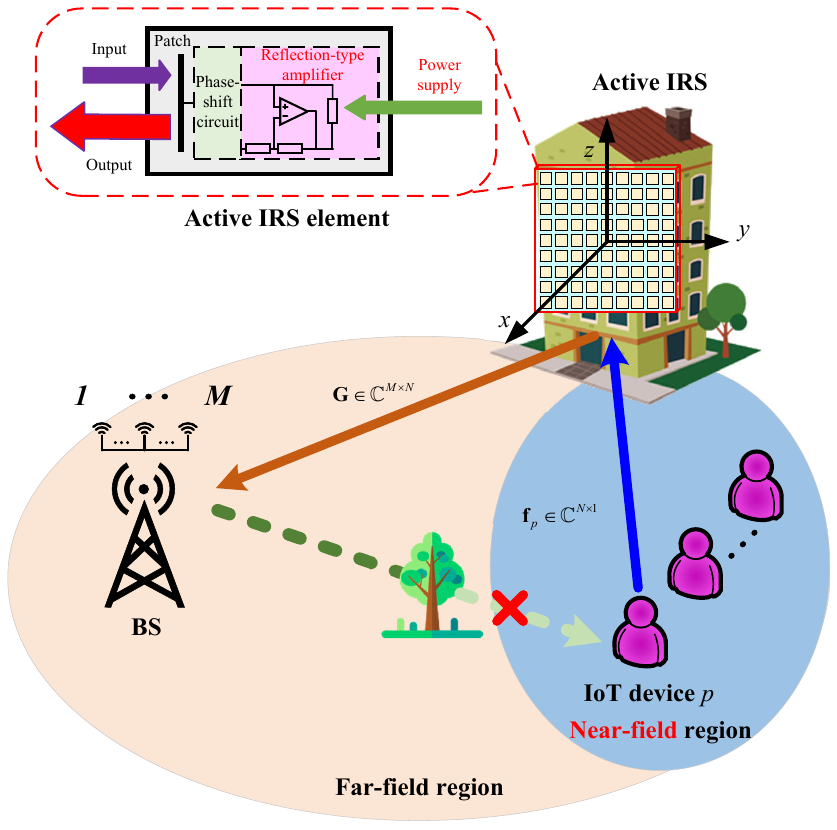}
\caption{A system model for active IRS-aided hybrid-field communication.}
\label{system-model}
\end{figure}

As shown in \ref{system-model}, an active IRS assisted hybrid-field IoT network is considered, in which the BS incorporates $M$ antennas and the active IRS integrates $N$ controllable reflective elements.
Owing to the fact that the direct communication link is heavily obscured by obstacles, $P$ single-antenna IoT devices establish a stable communication link with the BS with the help of an active IRS.
During the process of CE, orthogonal pilot sequences are commonly employed \cite{Chen2024Channel}. Consequently, for the purpose of uplink CE, we focus solely on an arbitrary user within the system.

Specifically, the signal sent out by the $p$-th IoT device is as follows
\begin{align}
s_p=\sqrt{\beta P_t}x_p,
\end{align}
where $x_p$ is the transmit symbol with unit energy, $P_t$ represents the total power of the IoT devices and active IRS, and $\beta$ denotes the PAF.

The received signal at active IRS is written as follows
\begin{align}\label{y-IRS}
\mathbf{y}_{\text{IRS}}=
\sqrt{\beta P_t}
\mathbf{\Theta}
\mathbf{f}_{p}x_{p}
+\mathbf{\Theta}
\mathbf{n}_i,
\end{align}
where $\mathbf{\Theta}\in\mathbb{C}^{N\times N}$ stands for the phase shift of the active IRS, and $\mathbf{f}_p\in\mathbb{C}^{N\times 1}$ denotes the channel between the $p$-th IoT devices and active IRS. 
$\mathbf{n}_i$ represents the additive white Gaussian noise (AWGN) at active IRS with $\mathbf{n}_i\in \mathbb{C}^{N\times 1}\sim \mathcal{CN}(\mathbf{0},\sigma_i^2\mathbf{I}_N)$.

Correspondingly, the received signal at BS can be modeled as follows
\begin{align}\label{y-BS-1}
\mathbf{y}&=
\sqrt{\beta P_t}
\mathbf{G}
\mathbf{\Theta}
\mathbf{f}_{p}x_{p}
+\mathbf{G}
\mathbf{\Theta}\mathbf{n}_i
+\mathbf{n}\nonumber\\
&=\sqrt{\beta P_t}
\underbrace{\mathbf{G}
\textit{diag}\{\mathbf{f}_{p}\}}_{\mathbf{H}_{p}}
\boldsymbol{\theta}x_{p}
+\mathbf{G}
\textit{diag}\{\mathbf{n}_i\}
\boldsymbol{\theta}
+\mathbf{n}\nonumber\\
&=\sqrt{\beta P_t}
\mathbf{H}_{p}
\boldsymbol{\theta}x_{p}
+\mathbf{G}
\mathbf{N}_{i}
\boldsymbol{\theta}
+\mathbf{n},
\end{align}
where $\mathbf{G}\in\mathbb{C}^{M\times N}$ represents the channels between active IRS and BS. 
$\mathbf{H}_{p}=\mathbf{G}\mathbf{F}_p
=\mathbf{G}\textit{diag}\{\mathbf{f}_{p}\}$ denotes the cascaded channel between the $p$-th IoT device and BS.
$\mathbf{n}$ is the AWGN at BS with $\mathbf{n}\in \mathbb{C}^{M\times 1}\sim \mathcal{CN}(\mathbf{0},\sigma^2\mathbf{I}_M)$.

For an active IRS, the reflection coefficient matrix $\mathbf{\Theta}$ is defined as $\textit{diag}\{\boldsymbol{\theta}\}$, where $\boldsymbol{\theta}\in\mathbb{C}^{N\times 1}$ can be decomposed into $\rho \boldsymbol{\tilde{\theta}}$. Here, $\rho= \|\boldsymbol{\theta}\|_2/\sqrt{N}$ and
$\|\boldsymbol{\tilde{\theta}}\|_2
=\sqrt{N}$.
On this basis, (\ref{y-BS-1}) can be reformulated as follows
\begin{align}\label{y-normalization-theta}
\mathbf{y}=\rho\sqrt{\beta P_t}
\mathbf{H}_{p}
\boldsymbol{\tilde{\theta}}x_p
+\rho\mathbf{G}
\mathbf{N}_{i}
\boldsymbol{\tilde{\theta}}
+\mathbf{n}.
\end{align}
Correspondingly, according to (\ref{y-IRS}), the reflected power generated by the active IRS is
\begin{align}\label{P-IRS}
P_{\text{IRS}}
&=\mathbb{E}\{\mathbf{y}_{\text{IRS}}^H
\mathbf{y}_{\text{IRS}}\}\\
&=\rho^2\beta P_t\|\mathbf{F}_p\boldsymbol{\tilde{\theta}}\|_2^2
+\rho^2\sigma_i^2
=(1-\beta)P_t.\nonumber
\end{align}
Based on (\ref{P-IRS}), $\rho$ can be mathematically expressed as
\begin{align}\label{rho}
\rho=\sqrt{\frac{(1-\beta)P_t}
{\beta P_t\|\mathbf{F}_p\boldsymbol{\tilde{\theta}}\|_2^2
+\sigma_i^2}}.
\end{align}

\vspace{-8pt}
\subsection{Channel Model}

\subsubsection{IRS-BS FF channel model}

\quad

Owing to the ``multiplicative fading'' effect induced by the IRS, it is preferable to deploy the active IRS either in proximity to the BS or adjacent to the IoT device \cite{Zhang2023Active}.
Given the relatively elevated positions of the BS and the IRS, the propagation conditions of the IRS-BS channel are less complex compared to those of the IoT device-IRS channel.
Consequently, the scenario of deploying IRS near IoT devices is focused on in this study, which enhances the capability to effectively control and optimize the channel between IoT devices and IRS.
Under this deployment assumption, due to the considerable distance between the IRS and the BS, the electromagnetic wave propagating between them exhibits a planar wavefront. 
Accordingly, the channel matrix $\mathbf{G}$ can be expressed as \cite{Wei2022Codebook}
\begin{align}
\mathbf{G}=
\eta_{G}
\mathbf{a}\big(\upsilon_{G_r},
\psi_{G_r}\big)
\mathbf{a}^T
\big(\upsilon_{G_t},
\psi_{G_t}\big),
\end{align}
where $\eta_G$ denotes the path gain, $\upsilon_{G_r}$ / $\psi_{G_r}$ and $\upsilon_{G_t}$ / $\psi_{G_t}$ represent the azimuth / elevation angle at BS and IRS, respectively.

In this study, both IRS and BS are considered as uniform planar array (UPA) configuration.
The steering vector $\mathbf{a}(\upsilon,\psi)$ for a BS with $M=M_y\times M_z$ antenna can be elaborated as follows
\begin{align}
\mathbf{a}(\upsilon,\psi)=&
\bigg[e^{-j\frac{2\pi}{\lambda}
\big(-m_y d_{yb}\sin \upsilon \cos\psi\big)}\bigg]\nonumber\\
&\otimes
\bigg[e^{-j\frac{2\pi}{\lambda}
\big(-m_z d_{zb}\sin \psi\big)}\bigg],
\end{align}
where $M_y=2\tilde{M}_y+1$, $M_z=2\tilde{M}_z+1$, and $\forall m_y\in\{-\tilde{M}_y,\ldots,\tilde{M}_y\}$, 
$\forall m_z\in\{-\tilde{M}_z,\ldots,\tilde{M}_z\}$. 
Moreover, $\lambda$ signifies the carrier wavelength, and the antenna spacing of BS's UPA array is configured as $d_{yb}$ and $d_{zb}$ in the two orthogonal dimensions, respectively.

\subsubsection{IoT device-IRS NF channel model}

\quad

Regarding the channel $\mathbf{f}_p$, given that the IRS is positioned in close proximity to the IoT device \cite{Shen2023Multi-Beam}, it is imperative to account for the spherical wavefront propagation characteristic when characterizing $\mathbf{f}_p$.

Specifically, the UPA of the active IRS is arranged within the $y-z$ plane and comprises $N = N_y \times N_z$ active elements.
Here, $N_y$ and $N_z$ are defined as $N_y=2\tilde{N}_y+1$ and $N_z=2\tilde{N}_z+1$, respectively.
The element spacings along the $y$-axis and $z$-axis directions are denoted as $d_{ya}$ and $d_{za}$, respectively.
The center of the IRS array is situated at the origin $(0,0,0)$ in the Cartesian coordinates.  
Consequently, the spatial coordinate of the $(n_y,n_z)$-th IRS element can be expressed as follows
\begin{align}
\mathbf{s}(n_y,n_z)=(0,~n_yd_{ya},
~n_zd_{za}),
\end{align}
where $\forall n_y\in \{-\tilde{N}_y,\ldots, \tilde{N}_y\}$ and $\forall n_z\in \{-\tilde{N}_z,\ldots, \tilde{N}_z\}$

As shown in Fig. \ref{3D-view}, the azimuth and elevation angles of the $p$-th user with respect to the UPA plane of the active IRS are $\alpha_p$ and $\beta_p$, respectively.
The distance between the $p$-th user and the center of the active IRS is $r_p$.
Then, the coordinate of the $p$-th user can be represented as
\begin{align}
\mathbf{u}_{p}=(u_{px},~u_{py},~u_{pz}).
\end{align}
Correspondingly, the polar coordinates of the $p$-th user with respect to the UPA plane of the active IRS are
\begin{align}\label{user-p}
\mathbf{u}_{p}&=(u_{px},~u_{py},~u_{pz})\nonumber\\
&=(r_p\cos \alpha_p\cos \beta_p,~
r_p\sin \alpha_p\cos \beta_p,~
r_p\sin \beta_p ).
\end{align}
In this case, the propagation distance between the $p$-th user and the $(n_y,n_z)$-th IRS element is
\begin{figure*}[b]
\centering	
\vspace*{1pt}
\hrulefill
\vspace*{1pt} 
\begin{equation}
\begin{split}
r_{p,(n_y,n_z)}
&=\|\mathbf{u}_{p}-\mathbf{s}(n_y,n_z)\|
=
\sqrt{
(r_p\cos \alpha_p\cos \beta_p)^2
+(r_p\sin \alpha_p\cos \beta_p-n_yd_{ya})^2
+(r_p\sin \beta_p -n_zd_{za})^2}\\
&=r_p\sqrt{1
+\frac{n_y^2d_{ya}^2}{r_p^2}
+\frac{n_z^2d_{za}^2}{r_p^2}
-\frac{2r_pn_yd_{ya}\sin \alpha_p\cos \beta_p}{r_p^2}
-\frac{2r_pn_zd_{za}\sin \beta_p}{r_p^2}}.
\end{split}
\end{equation}
\end{figure*}

Subsequently, by applying the following second-order Taylor expansion
\begin{align}
\sqrt{1+x}\approx1+\frac12x
-\frac18x^2+\mathcal{O}(x^2),
\end{align}
as well as omitting the bilinear term, we obtain
\begin{align}
r_{p,(n_y,n_z)}&=
r_p\nonumber\\
&\underbrace{-n_yd_{ya}\sin \alpha_p\cos \beta_p
+\frac{n_y^2d_{ya}^2(1-\sin ^2\alpha_p\cos ^2 \beta_p)}{2r_p}}_{r_{p,y},~\text{only depend on $n_y$}}\nonumber\\
&\underbrace{-n_zd_{za}\sin \beta_p
+\frac{n_z^2d_{za}^2\cos ^2\beta_p}{2r_p}}_{r_{p,z},~\text{only depend on $n_z$}}.
\end{align}
Upon eliminating the constant phase term $e^{-j\frac{2\pi}{\lambda}r_p}$, the phase of the array response vector (ARV) can be decomposed into two distinct components, denoted as $r_{p,y}$ and $r_{p,z}$, where $r_{p,y}$ is solely a function of $n_y$ and $r_{p,z}$ depends exclusively on $n_z$.

Then, the NF channel model between IoT devices and active IRS can be represented as
\begin{align}
\mathbf{f}_p=
\eta_{f,p}
\mathbf{b}
\big(\alpha_{p},\beta_{p},
r_{p}\big),
\end{align}
where the NF ARV for $\mathbf{f}_p$ can be given by
\begin{align}
\mathbf{b}(\alpha_p,\beta_p,r_p)=
\mathbf{c}_{y}(\alpha_p,\beta_p,r_p)
\otimes
\mathbf{c}_{z}(\beta_p,r_p).
\end{align}
Here, $\mathbf{c}_{y}(\alpha_p,\beta_p,r_p)$ and $\mathbf{c}_{z}(\beta_p,r_p)$ are respectively represented as follows
\begin{align}
&[\mathbf{c}_{y}
(\alpha_p,\beta_p,r_p)]_{n_y}\nonumber\\
&=
e^{-j\frac{2\pi}{\lambda}
\bigg(-n_yd_{ya}\sin \alpha_p\cos \beta_p
+\frac{n_y^2d_{ya}^2(1-\sin ^2\alpha_p\cos ^2 \beta_p)}{2r_p}\bigg)}
\end{align}
and
\begin{align}
[\mathbf{c}_{z}(\beta_p,r_p)]_{n_z}
=e^{-j\frac{2\pi}{\lambda}
\bigg(-n_zd_{za}\sin \beta_p
+\frac{n_z^2d_{za}^2\cos ^2\beta_p}{2r_p}\bigg)}.
\end{align}

\vspace{-8pt}
\subsection{Optimal PAF}

Unlike passive IRSs, which solely perform signal reflection, active IRSs possess dual functionalities of signal reflection and amplification \cite{Lin2023Enhanced-Rate}\cite{Shu2023Three}. 
Nevertheless, the signal amplification process in active IRSs inherently entails the introduction of additional amplification noise.
Consequently, the rational allocation of the total available power between IoT devices and active IRS has emerged as a pivotal challenge demanding immediate attention \cite{Wang2024Power}, with the primary aim of mitigating CE errors to the greatest extent possible.
Furthermore, the determination of the optimal PAF also facilitates the construction of datasets for the proposed CAEformer CE algorithm.

In order to estimate the uplink cascaded channel $\mathbf{H}_{p}$, the IoT device is required to send the known pilot signals to BS via the active IRS over $Q_1$ time slots, where $q_1\in\{1,2,\ldots,Q_1\}$.
According to (\ref{y-normalization-theta}), in the $q_1$-th time slot, the received pilot signal vector at BS is
\begin{align}
\mathbf{y}_{q_1}=\rho\sqrt{\beta P_t}
\mathbf{H}_{p}
\boldsymbol{\tilde{\theta}}x_{p,q_1}
+\rho\mathbf{G}
\mathbf{N}_{i,q_1}
\boldsymbol{\tilde{\theta}}
+\mathbf{n}_{q_1},
\end{align}
where $x_{p,q_1}$ represents the pilot signal sent by the IoT devices, $\mathbf{N}_{i,q_1}$ and $\mathbf{n}_{q_1}$ represent the noise at the active IRS and BS during the $q_1$-th time slot, respectively.

Stacking the $Q_1$ vectors at BS into a matrix form, as follows
\begin{align}\label{Y-Q1}
\mathbf{Y}&=
[\mathbf{y}_{1},\mathbf{y}_{2},
\ldots,\mathbf{y}_{Q_1}]
\nonumber\\
&=\rho\sqrt{\beta P_t}
\mathbf{H}_{p}
\mathbf{V}\mathbf{X}_{p}
+\rho\mathbf{G}
\mathbf{V} \mathbf{N}_{i} 
+\mathbf{N},
\end{align}
where 
$\mathbf{V}=[\boldsymbol{\tilde{\theta}}_{1}, \boldsymbol{\tilde{\theta}}_{2}, \ldots, \boldsymbol{\tilde{\theta}}_{Q_1}]
\in \mathbb{C}^{N\times Q_1}$, and $Q_1\geq N$ is used to guarantee the existence of the right pseudo-inverse of $\mathbf{V}$. 
Additionally, $\mathbf{x}_{p}=[x_{p,1}, x_{p,2}, \cdots, x_{p,Q_1}]^T\in \mathbb{C}^{Q_1\times 1}$ is the pilot sequence sent by IoT devices, and
$\mathbf{X}_{p}=\textit{diag}\{\mathbf{x}_p\}\in \mathbb{C}^{Q_1\times Q_1}$.
$\mathbf{N}_{i}\in \mathbb{C}^{Q_1\times Q_1}$
and
$\mathbf{N}
=\big[
\mathbf{n}_{1},
\mathbf{n}_{2},
\ldots,
\mathbf{n}_{Q_1}
\big]\in\mathbb{C}^{M\times Q_1}$ are both noise terms.

By applying the vectorization operation, denoted as $\textit{vec}(\cdot)$, to both sides of the preceding equation (\ref{Y-Q1}), and leveraging the property $\textit{vec}(\mathbf{A}\mathbf{C})
=(\mathbf{C}^T\otimes\mathbf{I})
\textit{vec}(\mathbf{A})$, the following expression is obtained:
\begin{align}\label{vec-Y}
\textit{vec}(\mathbf{Y})
=&\rho\sqrt{\beta P_t}
(\underbrace{(\mathbf{V}
\mathbf{X}_p)^T\otimes\mathbf{I}_M}_{\mathbf{A}})
\textit{vec}(\mathbf{H}_{p})\nonumber\\
&
+\rho\textit{vec}
(\mathbf{G}
\mathbf{V}
\mathbf{N}_{i})
+\textit{vec}(\mathbf{N}),
\end{align}
therefore, the LS estimation of $\mathbf{H}_{p}$ can be obtained through (\ref{vec-Y}) as follows
\begin{align}
\textit{vec}(\widehat{\mathbf{H}}_{p})
=\frac{\mathbf{A}^{-1}
\textit{vec}(\mathbf{Y})}
{\rho\sqrt{\beta P_t}}.
\end{align}
The MSE associated with the estimation of $\mathbf{H}_{p}$ is given by the following expression
\begin{align}\label{MSE-rho}
\varepsilon
&=\frac{1}{N}\mathbb{E}\{\|\textit{vec}
(\widehat{\mathbf{H}}_{p}
)
-\textit{vec}(\mathbf{H}_{p})\|_F^2\}
\nonumber\\
&=\frac{1}{N}\mathbb{E}
\bigg\{\bigg\|\frac{\mathbf{A}^{-1}
\textit{vec}(\mathbf{Y})}
{\rho\sqrt{\beta P_t}}
-\textit{vec}(\mathbf{H}_{p})\bigg\|_F^2\bigg\}
\nonumber\\
&=\frac{
\rho^2\sigma_i^2
\|\mathbf{B}\|^2_F
+\sigma^2\|\mathbf{A}^{-1}\|^2_F}
{N\rho^2\beta P_t}
\end{align}
where $\mathbf{B}=\mathbf{A}^{-1}
(\mathbf{I}_{N}\otimes
\mathbf{G}
\mathbf{V})$.

To facilitate the derivation of the optimal PAF, substituting Eq. (\ref{rho}) into Eq. (\ref{MSE-rho}) yields the MSE function with respect to $\beta$, as follows
\begin{align}\label{varepsilon-beta}
\varepsilon(\beta)
=\frac{a_1\beta+a_2}{b\beta^2-b\beta},
\end{align}
where
\begin{align}
a_1&=
\sigma^2 P_t
\|\mathbf{F}_p\boldsymbol{\tilde{\theta}}\|_2^2
\cdot \|\mathbf{A}^{-1}\|^2_F
-\sigma_i^2 P_t
\|\mathbf{B}\|^2_F
\nonumber\\
a_2&=\sigma_i^2 P_t\|\mathbf{B}\|^2_F
+\sigma^2\sigma_i^2\|\mathbf{A}^{-1}\|^2_F
\nonumber\\
b&=-NP_t^2.
\end{align}
Consequently, the optimization problem regarding $\beta$ can be constructed as follows
\begin{align}
\underset{\beta}{\mathrm{min}} \quad &\varepsilon(\beta) \label{Problem}\\
\mathrm{s.t.} \quad &0<\beta<1\tag{\ref{Problem}{a}} .
\end{align}
The derivative of (\ref{varepsilon-beta}) with respect to $\beta$ is
\begin{align}
\varepsilon^{\prime}(\beta)
=\frac{-a_1b\beta^2
-2a_2b\beta+a_2b}
{(b\beta^2-b\beta)^2}.
\end{align}
Let the above equation be equal to 0, namely
\begin{align}
-a_1b\beta^2
-2a_2b\beta+a_2b=0,
\end{align}
subsequently, we can obtain
\begin{align}\label{beta-opt}
\beta^{\text{opt}}
=\underset{\beta\in S_1}{\arg\min}\quad(\ref{varepsilon-beta}),
\end{align}
where $S_1=\{\beta_1, \beta_2\}$ with
\begin{align}\label{optimalbeta}
\beta_{1} 
=\frac{a_2+\sqrt{a_2^2+a_1a_2}}{-a_1}, 
~\beta_{2}
=\frac{a_2-\sqrt{a_2^2+a_1a_2}}{-a_1}.
\end{align}

\section{Proposed block-based far-field channel model}


In this section, to begin with, a block-based FF channel model is proposed to approximate NF channel model.
Subsequently, the channel approximation error and CEE are derived separately.
Finally, the relationship between the total error and the number of sub-block is established.

\vspace{-8pt}
\subsection{Proposed block-based FF channel model}

\begin{figure}[h]
\centering
\includegraphics[width=3.4in]
{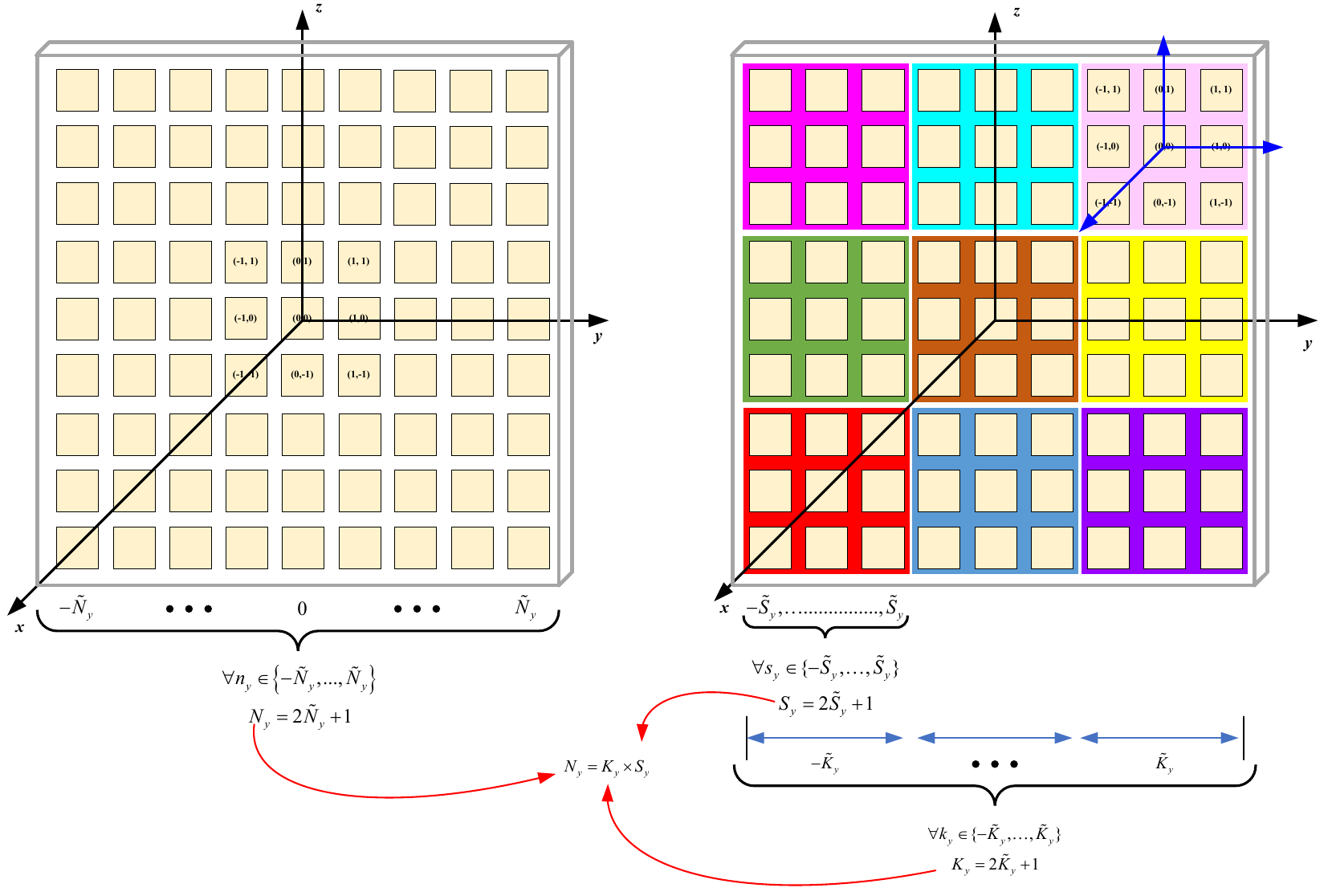}
\caption{Non-block-based active IRS and block-based active IRS.}
\label{Visio-Block}
\end{figure}

One of the challenges in CE for the cascaded channel of IRS-assisted wireless networks lies in the high dimensionality of the cascaded channel \cite{Guan2022Anchor-Assisted}\cite{Liu2020Uplink-Aided}. Moreover, the presence of numerous channel characterization parameters in NF channels exacerbates the overhead associated with CE.
To mitigate the CE overhead, inspired by \cite{Cui2024Near-Field}, as illustrated in Fig.~\ref{Visio-Block}, the IRS is partitioned into blocks. Consequently, the original NF channel between the IoT device and the IRS can be approximated by a block-based FF channel model.

To elaborate, as described in Fig.~\ref{Visio-Block}, the entire IRS is divided into $K_y$ and $K_z$ sub blocks on the $y$-axis and $z$-axis, respectively, where $K_y=2\tilde{K}_y+1$ and $K_z=2\tilde{K}_z+1$.
In other words, in the $y$-axis and $z$-axis directions, each sub block contains $S_y=\frac{N_y}{K_y}$ and $S_z=\frac{N_z}{K_z}$ elements, respectively, where $S_y=2\tilde{S}_y+1$ and $S_z=2\tilde{S}_z+1$.
Therefore, the entire IRS is divided into $K=K_y \times K_z$ sub blocks, with the number of components in each sub block being $S=S_y \times S_z$, i.e. $N_y=K_y\times S_y$ and $N_z=K_z\times S_z$.

\begin{figure}[h]
\centering
\includegraphics[width=3.4in]
{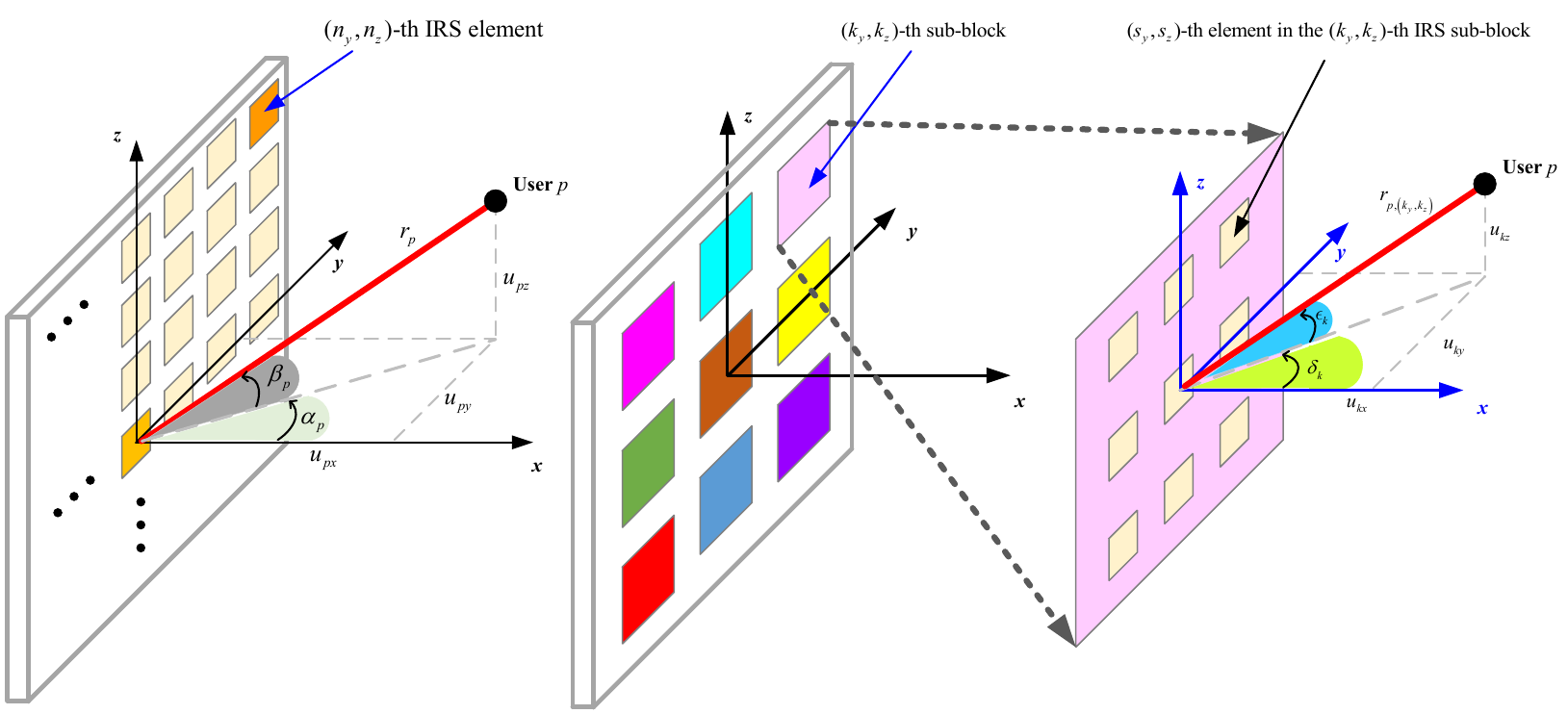}
\caption{3D view of block-based active IRS.}
\label{3D-view}
\end{figure}

Consequently, as depicted in Fig. \ref{3D-view}, the coordinate of the center of the $(k_y,k_z)$-th IRS sub-block is 
\begin{align}
\mathbf{s}_{b}(k_y,k_z)=(0,~k_yS_yd_{ya},
~k_zS_zd_{za}),
\end{align}
where $\forall k_y\in \{-\tilde{K}_y,\ldots, \tilde{K}_y\}$ and $\forall k_z\in \{-\tilde{K}_z,\ldots, \tilde{K}_z\}$.
Then, the distance of the center of the $(k_y,k_z)$-th IRS sub-block and the $p$-th user is
\begin{align}
&r_{p,(k_y,k_z)}
=\|\mathbf{u}_{p}-\mathbf{s}_b(k_y,k_z)\|
\nonumber\\
&=\bigg(
(r_p\cos \alpha_p\cos \beta_p)^2
+(r_p\sin \alpha_p\cos \beta_p-k_yS_yd_{ya})^2\nonumber\\
&+(r_p\sin \beta_p -k_zS_zd_{za})^2
\bigg)^{\frac{1}{2}}.
\end{align}

Assuming that the azimuth and elevation angles of the $p$-th user with respect to the $(k_y,k_z)$-th IRS sub-block are $\delta_k$ and $\epsilon_k$, respectively.
Therefore, similar to (\ref{user-p}), the polar coordinates of the $p$-th user with respect to the $(k_y,k_z)$-th IRS sub-block can be represented as
\begin{align}\label{u-p-k}
\mathbf{u}_{p,(k_y,k_z)}&
=(u_{kx},~u_{ky},~u_{kz})\nonumber\\
&=(r_{p,(k_y,k_z)}\cos \delta_k\cos \epsilon_k,~
r_{p,(k_y,k_z)}\sin \delta_k\cos \epsilon_k,~\nonumber\\
&~~~~r_{p,(k_y,k_z)}\sin \epsilon_k).
\end{align}
Subsequently, the coordinate of the $(s_y,s_z)$-th element in the $(k_y,k_z)$-th IRS sub-block is
\begin{align}\label{s-e}
\mathbf{s}_e(s_y,s_z)=(0,~s_yd_{ya},
~s_zd_{za}),
\end{align}
where $\forall s_y\in \{-\tilde{S}_y,\ldots, \tilde{S}_y\}$ and $\forall s_z\in \{-\tilde{S}_z,\ldots, \tilde{S}_z\}$.
Combining Eqs. (\ref{u-p-k}) and (\ref{s-e}), the distance of the $(s_y,s_z)$-th element in the $(k_y,k_z)$-th IRS sub-block and the $p$-th user is
\begin{align}\label{r-p-k-s}
r_{p,(k_y,k_z)}^{(s_y,s_z)}
&=\|\mathbf{u}_{p,(k_y,k_z)}
-\mathbf{s}_e(s_y,s_z)\|
\\
&\overset{(a)}{\approx} 
r_{p,(k_y,k_z)}
-s_yd_{ya}\sin \delta_k\cos \epsilon_k
-s_zd_{za}\sin \epsilon_k,\nonumber
\end{align}
here, the approximation (a) holds because of Taylor expansion.

Therefore, the channel between user $p$ and the $k$-th sub-block of the active IRS can be modeled as
\begin{align}
\mathbf{f}_{p, k}&=\mathbf{f}_{p, (k_y,k_z)}
\in\mathbb{C}^{S\times 1}
\\
&=\bigg[e^{-j\frac{2\pi}{\lambda}
\big(-s_y d_{ya}\delta_k\cos \epsilon_k\big)}\bigg]
\otimes
\bigg[e^{-j\frac{2\pi}{\lambda}
\big(-s_z d_{za}\sin \epsilon_k\big)}\bigg],\nonumber
\end{align}
then, the complex NF channel between the $p$-th IoT device and the active IRS can be approximated as a block-based FF channel, as follows:
\begin{align}
\tilde{\mathbf{f}}_{p}=&
\bigg[
\mathbf{f}_{p, (-\tilde{K}_y,-\tilde{K}_z)}^T,
\ldots,
\mathbf{f}_{p,(-\tilde{K}_y,\tilde{K}_z)}^T,
\ldots,
\nonumber\\
&
\mathbf{f}_{p, (\tilde{K}_y,-\tilde{K}_z)}^T,
\ldots,
\mathbf{f}_{p, (\tilde{K}_y,\tilde{K}_z)}^T
\bigg]^T
\in\mathbb{C}^{KS\times 1}.
\end{align}
Table \ref{tab-block} compares the symbols and dimensions of different channels between non-block-based active IRS and block-based active IRS.
\begin{table*}[h]
\begin{center}
\renewcommand{\arraystretch}{1.8}
\tabcolsep=0.2cm
\caption{Notation comparison.}
\label{tab-block}
\begin{tabular}{c c c c}
\hline
\textbf{Meaning of notations} & \textbf{Non-block-based IRS} & \textbf{The $k$-th sub-block-based IRS}& \textbf{Block-based IRS}\\  
\hline
IoT device $\rightarrow$ active IRS channel& $\mathbf{f}_{p}\in\mathbb{C}^{N\times 1}$ & $\mathbf{f}_{p, k}\in\mathbb{C}^{S\times 1}$
&$\tilde{\mathbf{f}}_{p}=\big[
\mathbf{f}_{p, 1}^T,
\mathbf{f}_{p, 2}^T,\ldots
\mathbf{f}_{p, K}^T
\big]^T\in\mathbb{C}^{KS\times 1}$\\
Active IRS $\rightarrow$ BS channel & $\mathbf{G}\in\mathbb{C}^{M\times N}$  &$\mathbf{G}_{k}\in\mathbb{C}^{M\times S}$
&$
\widetilde{\mathbf{G}}=
\big[\mathbf{G}_1,
\mathbf{G}_2,
\ldots,
\mathbf{G}_K
\big]\in \mathbb{C}^{M\times KS}$\\ 
Cascaded channel&$\mathbf{H}_{p}\in\mathbb{C}^{M\times N}$ & $\mathbf{H}_{p, k}\in\mathbb{C}^{M\times S}$
&$\widetilde{\mathbf{H}}_{p}
=\big[\mathbf{H}_{p,1},
\mathbf{H}_{p,2},\ldots,
\mathbf{H}_{p,K}\big]\in \mathbb{C}^{M\times KS}$\\
\hline
\end{tabular}
\end{center}
\end{table*}

\begin{figure*}[!t]
\centering
\subfloat[Near-field]
{\includegraphics[width=0.25\textwidth]
{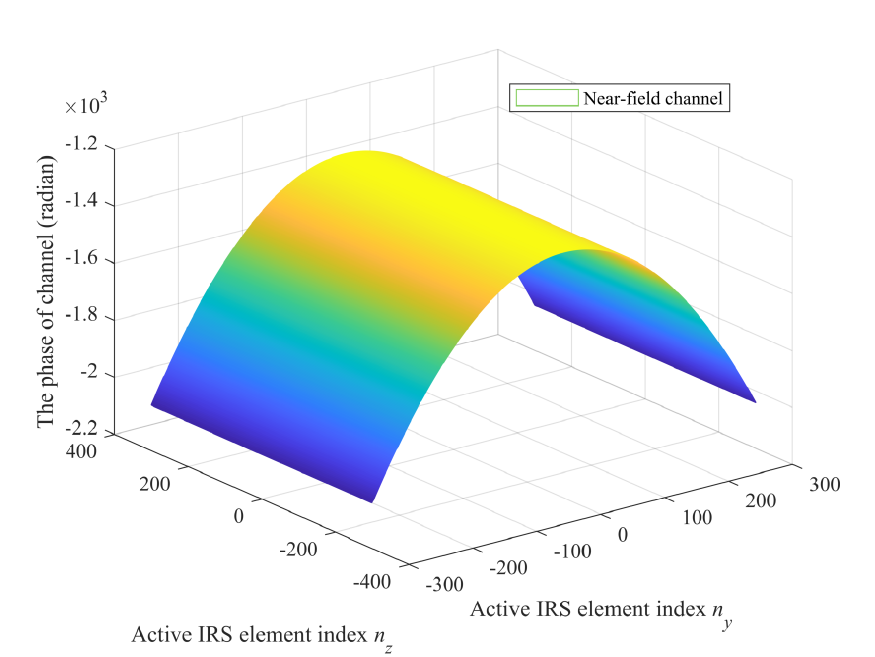}}
\subfloat[Block-based far-field $K=9$]
{\includegraphics[width=0.25\textwidth]
{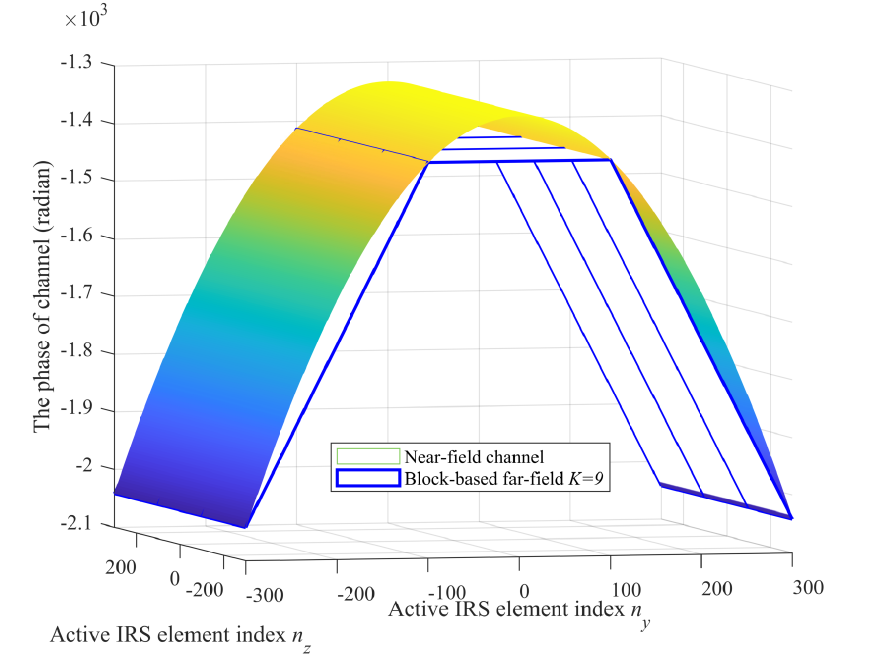}}
\subfloat[Block-based far-field $K=36$]
{\includegraphics[width=0.25\textwidth]
{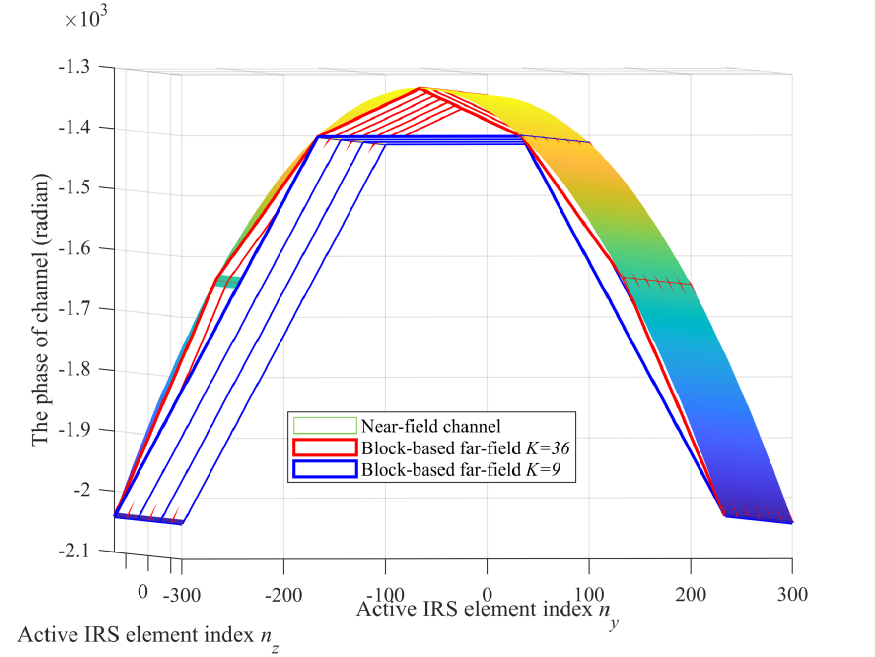}}
\subfloat[Far-field]
{\includegraphics[width=0.25\textwidth]
{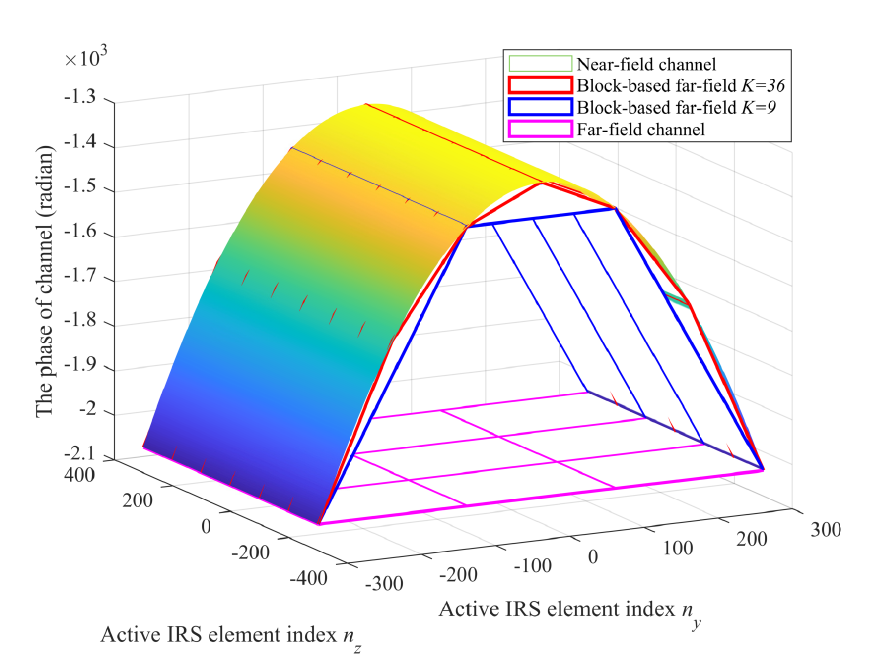}}
\caption{The phase of channel versus the active IRS element index. As the number of sub-block $K$ increases, the block-based far-field channel model can well approximate the near-field channel model.}
\label{3D-channel-phase}
\end{figure*}

\vspace{-10pt}
\subsection{Channel Approximation Error}

Eq. (\ref{Y-Q1}) can be rephrased as
\begin{align}
\mathbf{Y}=&
\rho\sqrt{\beta P_t}
\mathbf{H}_{p}
\mathbf{V}\mathbf{X}_{p}
+\rho\mathbf{G}
\mathbf{V} \mathbf{N}_{i} 
+\mathbf{N}
\nonumber\\
=&
\rho\sqrt{\beta P_t}
\bigg(\widetilde{\mathbf{H}}_{p}
+\underbrace{\mathbf{H}_{p}
-\widetilde{\mathbf{H}}_{p}}_{\Delta \mathbf{H}_{p},~~\text{approximation error}}\bigg)
\mathbf{V}\mathbf{X}_{p}
\nonumber\\
&+\underbrace{\rho\mathbf{G}
\mathbf{V} \mathbf{N}_{i} 
+\mathbf{N}}_{\text{estimation error}},
\end{align}
where $\Delta \mathbf{H}_{p}$ can be expressed as
\begin{align}
\Delta \mathbf{H}_{p}=
\mathbf{H}_{p}
-\widetilde{\mathbf{H}}_{p}
&=\mathbf{G}
\bigg(\text{diag}\{\mathbf{f}_{p}\}
-\text{diag}\{\widetilde{\mathbf{f}}_{p}\}\bigg)
\nonumber\\
&=\mathbf{G}
\text{diag}\{\mathbf{f}_{ p}
-\widetilde{\mathbf{f}}_{p}\}
=\mathbf{G}
\text{diag}\{\Delta \mathbf{f}_p\}.
\end{align}
As can be seen from the above equation, the channel approximation error is related to $\Delta \mathbf{f}_p$.

Before blocking, the IoT device-IRS NF channel is related to the distance from the $p$-th IoT device to each IRS element, namely
\begin{align}
\mathbf{f}_p(n_y,n_z)=
e^{-j\frac{2\pi}{\lambda}
r_{p,(n_y,n_z)}},
\end{align}
there are $N$ different distances in total.
Therefore, the entire actual NF channel $\mathbf{f}_p$ can be represented as
\begin{align}
\mathbf{f}_p=\bigg[
&e^{-j\frac{2\pi}{\lambda}
r_{p,(-\tilde{N}_y,-\tilde{N}_z)}},\ldots,
e^{-j\frac{2\pi}{\lambda}
r_{p,(-\tilde{N}_y,\tilde{N}_z)}},\ldots,\nonumber\\
&e^{-j\frac{2\pi}{\lambda}
r_{p,(\tilde{N}_y,-\tilde{N}_z)}},\ldots,
e^{-j\frac{2\pi}{\lambda}
r_{p,(\tilde{N}_y,\tilde{N}_z)}}
\bigg]^T.
\end{align}

After blocking, the distance from the $p$-th user to the center of each sub-block is used as the distance of all elements in that sub-block, as follows
\begin{align}
\widetilde{\mathbf{f}}_{p}
=\bigg[
&\underbrace{e^{-j\frac{2\pi}{\lambda}
r_{p,(-\tilde{K}_y,-\tilde{K}_z)}},\ldots,
e^{-j\frac{2\pi}{\lambda}
r_{p,(-\tilde{K}_y,-\tilde{K}_z)}},}_{\text{S elements}}
\ldots,
\nonumber\\
&\underbrace{e^{-j\frac{2\pi}{\lambda}
r_{p,(\tilde{K}_y,\tilde{K}_z)}},\ldots,
e^{-j\frac{2\pi}{\lambda}
r_{p,(\tilde{K}_y,\tilde{K}_z)}}}_{\text{S elements}}
\bigg]^T.
\end{align}

According to the NF channel characteristics mentioned in 
\cite{Liu2023Near-Field}, the amplitude error can be ignored. Therefore, the channel approximation error for each sub-block is as follows
\begin{align}\label{varepsilon-k}
\varepsilon_{k_y,k_z}=
\sum_{s_y=-\tilde{S}_y}^{\tilde{S}_y}
\sum_{s_z=-\tilde{S}_z}^{\tilde{S}_z}
|e^{-j\frac{2\pi}{\lambda}
r_{p,(k_y,k_z)}^{(s_y,s_z)}}
-e^{-j\frac{2\pi}{\lambda}
r_{p,(k_y,k_z)}}|^2,
\end{align}
then, the channel approximation error of all sub blocks can be expressed as
\begin{align}\label{appro-actual}
\varepsilon_{\text{approximation}}
=&
\sum_{k_y=-\tilde{K}_y}^{\tilde{K}_y}
\sum_{k_z=-\tilde{K}_z}^{\tilde{K}_z}
\varepsilon_{k_y,k_z}.
\end{align}

Let's define
\begin{align}
\Delta r=
r_{p,(k_y,k_z)}^{(s_y,s_z)}
-r_{p,(k_y,k_z)},
\end{align}
based on this, we have
\begin{align}
&\bigg|e^{-j\frac{2\pi}{\lambda}
r_{p,(k_y,k_z)}^{(s_y,s_z)}}
-e^{-j\frac{2\pi}{\lambda}
r_{p,(k_y,k_z)}}\bigg|^2
\nonumber\\
=&
\bigg(e^{-j\frac{2\pi}{\lambda}\Delta r}-1\bigg)
\bigg(e^{j\frac{2\pi}{\lambda}\Delta r}-1\bigg)
\nonumber\\
\overset{(b)}{=}&
2-2\cos\bigg(\frac{2\pi}{\lambda}\Delta r\bigg)
\overset{(c)}{\approx}
\bigg(\frac{2\pi}{\lambda}\Delta r\bigg)^2,
\end{align}
where (b) holds because of $e^{ix}=\cos x+i\sin x$,
and (c) holds because of 
$\cos x\approx1-\frac{1}{2}x^2$.

As a result, (\ref{varepsilon-k}) can be rephrased as
\begin{align}\label{k-Delta-r}
\varepsilon_{k_y,k_z}=
\sum_{s_y=-\tilde{S}_y}^{\tilde{S}_y}
\sum_{s_z=-\tilde{S}_z}^{\tilde{S}_z}
\bigg(\frac{2\pi}{\lambda}\Delta r\bigg)^2.
\end{align}
According to Eq. (\ref{r-p-k-s}), it can be concluded that:
\begin{align}
\Delta r&=
r_{p,(k_y,k_z)}^{(s_y,s_z)}
-r_{p,(k_y,k_z)}\nonumber\\
&\approx
-s_yd_{ya}\sin \delta_k\cos \epsilon_k
-s_zd_{za}\sin \epsilon_k\nonumber\\
&=a_ys_y+a_zs_z,
\end{align}
where $a_y=-d_{ya}\sin \delta_k\cos \epsilon_k$ and $a_z=-d_{za}\sin \epsilon_k$.
By substituting the above equation into Eq. (\ref{k-Delta-r}), we can obtain
\begin{align}
\varepsilon_{k_y,k_z}=
\bigg(\frac{2\pi}{\lambda}\bigg)^2
\sum_{s_y=-\tilde{S}_y}^{\tilde{S}_y}
\sum_{s_z=-\tilde{S}_z}^{\tilde{S}_z}
(a_y^2s_y^2+a_z^2s_z^2+2a_ya_zs_y s_z),
\end{align}
Due to $s_y\in \{-\tilde{S}_y,\ldots, \tilde{S}_y\}$ and $s_y\in \{-\frac{S_y-1}{2},\ldots, \frac{S_y-1}{2}\}$, we can deduce
\begin{align}
&\sum_ {s_y=-\tilde{S}_y}^{\tilde{S}_y}
\sum_{s_z=-\tilde{S}_z}^{\tilde{S}_z}
2a_ya_zs_y s_z=0,\nonumber\\
&\sum_{s_y=-\tilde{S}_y}^{\tilde{S}_y}
s_y^2
=\frac{S_y(S_y^2-1)}{12},\nonumber\\
&\sum_{s_z=-\tilde{S}_z}^{\tilde{S}_z}
s_z^2
=\frac{S_z(S_z^2-1)}{12},
\end{align}
thus, the channel approximation error of each sub-block is a function of $S_y$ and $S_z$, as follows
\begin{align}
\varepsilon_{k_y,k_z}&=
\bigg(\frac{2\pi}{\lambda}\bigg)^2
\bigg(
\frac{a_y^2}{12} S_zS_y(S_y^2-1)
+\frac{a_z^2}{12} S_yS_z(S_z^2-1)
\bigg).
\end{align}
Assuming equal partitioning in both the $y$-axis and $z$-axis directions, i.e, $S_z=\sqrt{S}$ and $S_y=\sqrt{S}$, we get
\begin{align}
\varepsilon_{k_y,k_z}=
C_1S(S-1),
\end{align}
where $C_1=\bigg(
\frac{2\pi}{\lambda}\bigg)^2
\bigg(
\frac{a_y^2}{12}
+\frac{a_z^2}{12}
\bigg)$.
In the end, the channel approximation error of the entire $K$ block is
\begin{align}\label{appro-close-form}
\varepsilon_{\text{approximation}}
(K)
&=
\frac{C_2N^2}{K^2}-\frac{C_2N}{K},
\end{align}
where $C_2=\sum_{k_y=-\tilde{K}_y}^{\tilde{K}_y}
\sum_{k_z=-\tilde{K}_z}^{\tilde{K}_z}
C_1$.

To validate the fidelity of the proposed block-based FF channel model, Fig. \ref{3D-channel-phase}
depicts the channel phase of the NF, FF, and block-based FF channel models as a function of the index of the active IRS elements.
As the number of sub blocks increases, the channel phase of the block-based FF channel model progressively converges to that of the true NF channel.
In essence, this channel model serves as a block linear approximation of the intricate NF channel model, wherein the phase exhibits local linear variation characteristics within each sub-block.

\subsection{Channel Estimation Error}

\begin{figure}[h]
\centering
\includegraphics[width=3.4in]
{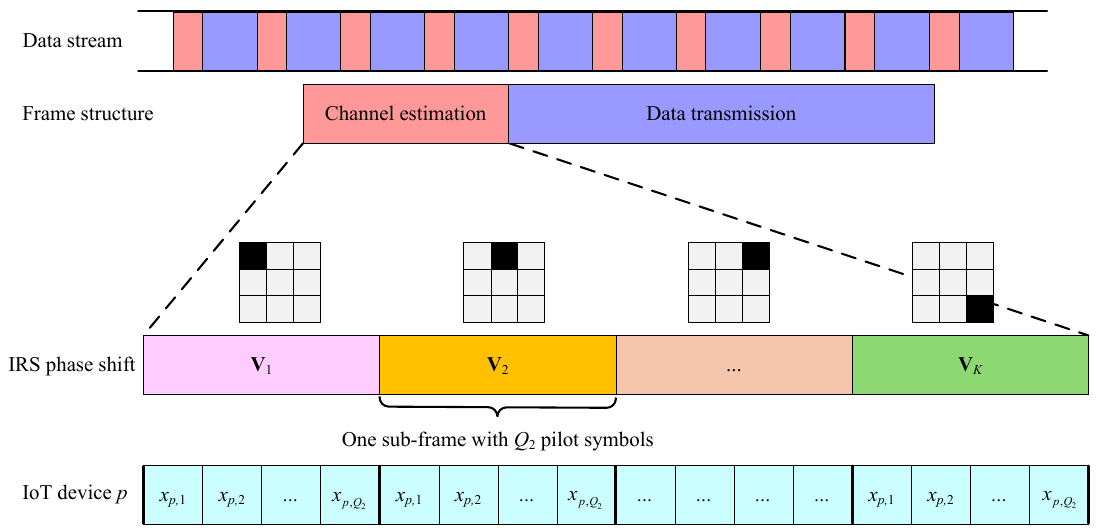}
\caption{Pilot pattern for the considered block-based active IRS-aided hybrid-field communication system.}
\label{Pilot-pattern}
\end{figure}

In order to estimate the channel after partitioning, a CE protocol for the considered block-based active IRS-assisted hybrid-field communication system is designed.
As illustrated in Fig. \ref{Pilot-pattern}, the CE phase comprises $K$ sub-frames, with each sub-frame containing $Q_2$ pilot symbols.
For the IRS, the phase-shift matrix remains constant within a single sub-frame and transitions to distinct phase-shift matrices across different sub-frames \cite{Liu2023Deep}.

In the $q_2$-th, $q_2\in\{1,2,\ldots,Q_2\}$, received pilot signal at BS in the $k$-th sub-frame is
\begin{align}
\mathbf{y}_{k,q_2}=\rho\sqrt{\beta P_t/K}
\mathbf{H}_{p,k}
\boldsymbol{\tilde{\theta}}_{k}x_{p,q_2}
+\rho\mathbf{G}_{k}
\mathbf{N}_{i,k,q_2}
\boldsymbol{\tilde{\theta}}_{k}
+\mathbf{n}_{k,q_2}.
\end{align}
It is worth noting that the power required to estimate the channel involving each sub-block is $P_t/K$, provided that the total power is constant. In addition, $\boldsymbol{\tilde{\theta}}_{k}
\in\mathbb{C}^{S\times 1}$ is the phase shift in the $k$-th sub-frame, $\mathbf{n}_{k,q_2}$ and $\mathbf{N}_{i,k,q_2}$ are the noise at the BS and active IRS in the $k$-th sub-frame, respectively.

Similar to Eq. (\ref{Y-Q1}), we have
\begin{align}\label{Y-k-sub}
\mathbf{Y}_{k}&=
[\mathbf{y}_{k,1},\mathbf{y}_{k,2},
\ldots,\mathbf{y}_{k,Q_2}]
\nonumber\\
&=\rho\sqrt{\beta P_t/K}
\mathbf{H}_{p, k}
\mathbf{V}_{k}\mathbf{X}_{p,k}
+\rho\mathbf{G}_{k}
\mathbf{V}_{k} \mathbf{N}_{i,k} 
+\mathbf{N}_k,
\end{align}
where $\mathbf{x}_{p,k}=[x_{p,1}, x_{p,2}, \cdots, x_{p,Q_2}]^T\in \mathbb{C}^{Q_2\times 1}$, $\mathbf{X}_{p,k}=
\textit{diag}\{\mathbf{x}_{p,k}\}\in \mathbb{C}^{Q_2\times Q_2}$, $\mathbf{V}_k=[
\boldsymbol{\tilde{\theta}}_{k,1}, \boldsymbol{\tilde{\theta}}_{k,2}, \ldots, \boldsymbol{\tilde{\theta}}_{k,Q_2}]
\in \mathbb{C}^{S\times Q_2}$ and $\mathbf{N}_k
=\big[
\mathbf{n}_{k,1},
\mathbf{n}_{k,2},
\ldots,
\mathbf{n}_{k,Q_2}
\big]\in\mathbb{C}^{M\times Q_2}$.

Similar to the derivation process of (\ref{vec-Y}), Eq. (\ref{Y-k-sub}) can be rewritten as
\begin{align}
\textit{vec}(\mathbf{Y}_{k})
=&\rho\sqrt{\beta P_t/K}
(\underbrace{(\mathbf{V}_{k}
\mathbf{X}_{p,k})^T\otimes\mathbf{I}_M}_{\mathbf{A}_k})
\textit{vec}(\mathbf{H}_{p, k})\nonumber\\
&
+\rho\textit{vec}
(\mathbf{G}_{k}
\mathbf{V}_{k}
\mathbf{N}_{i,k})
+\textit{vec}(\mathbf{N}_k),
\end{align}
which subsequently yields the LS estimation of $\mathbf{H}_{p, k}$ as follows
\begin{align}
\textit{vec}(\widehat{\mathbf{H}}_{p, k})
=\frac{\sqrt{K}\mathbf{A}_k^{-1}
\textit{vec}(\mathbf{Y}_{k})}
{\rho\sqrt{\beta P_t}}.
\end{align}
The estimated MSE for each sub block cascaded channel is
\begin{align}
\varepsilon_k
=\frac{1}{S}\mathbb{E}\{\|\textit{vec}
(\widehat{\mathbf{H}}_{p, k}
)
-\textit{vec}(\mathbf{H}_{p, k})\|_F^2\}
=\frac{C_3K^2}{N},
\end{align}
where
\begin{align}
C_3=\frac{1}{\rho^2\beta P_t}
(\rho^2\sigma_i^2
\|\mathbf{A}_k^{-1}
(\mathbf{I}_S\otimes
\mathbf{G}_{k}
\mathbf{V}_{k}
)\|^2_F
+\sigma^2\|\mathbf{A}_k^{-1}\|^2_F).
\end{align}
In the end, The CEE for all $K$ sub block channels is
\begin{align}\label{estimation-K}
\varepsilon_{\text{estimation}}(K)
=\sum_{k=1}^{K}\varepsilon_k
=\frac{C_3K^3}{N}.
\end{align}

\subsection{Total Error versus the Optimal Number of Sub-Block $K^*$}

Combining the aforementioned channel approximation error and CEE, the total error can be expressed as follows
\begin{align}\label{varepsilon-K}
\varepsilon_{\text{total}}(K)&=
\varepsilon_{\text{approximation}}(K)
+\varepsilon_{\text{estimation}}(K)
\nonumber\\
&=\frac{C_3K^5-C_2N^2K+C_2N^3}{NK^2},
\end{align}
based on this, the optimization objective of $K$ can be casted as
\begin{align}
\underset{K}{\mathrm{min}} \quad &
\varepsilon_{\text{total}}(K) 
\label{ProblemK}\\
\mathrm{s.t.} \quad &1\leq K\leq N
\tag{\ref{ProblemK}{a}} .
\end{align}
The derivative of (\ref{varepsilon-K}) with respect to $K$ yields
\begin{align}\label{varepsilon-prime}
\varepsilon^{\prime}_{\text{total}}
(K)
=\frac{3C_3K^5+C_2N^2K-2C_2N^3}{NK^3}
\end{align}
Letting (\ref{varepsilon-prime}) equal 0, namely
\begin{align}
3C_3K^5+C_2N^2K-2C_2N^3=0,
\end{align}
here, the optimal $K$ is shown to be one root of a fifth-order polynomial. 
The solution of the optimal $K$ consists of two steps: firstly, one candidate root, i.e., $K_1$, is obtained using the Newton-Raphson algorithm \cite{Shu2024Intelligent}, and the order of the polynomial is reduced from five to four, and secondly, the remaining four feasible solutions, i.e., $K_2$, $K_3$, $K_4$, and $K_5$, can be obtained by the Ferrari's method \cite{Wang2024Power}.

In the end,  the optimal solution is given as follows
\begin{align}
K^{\text{opt}}
=\underset{K\in S_2}{\arg\min}\quad(\ref{varepsilon-K}),
\end{align}
where 
\begin{align}
S_2=\{K_1, K_2,K_3, K_4, K_5\}.
\end{align}

\section{CRLB}

To evaluate the performance of the subsequently proposed CAEformer CE algorithm, CRLB is derived in this section. 
Firstly, according to Eq. (\ref{vec-Y}), we have
\begin{align}\label{y-hx-n}
\mathbf{y}_p=\mathbf{A}_p\mathbf{h}_p+\mathbf{n}_p,
\end{align}
where $\mathbf{y}_p=\textit{vec}(\mathbf{Y})$, $\mathbf{A}_p=\rho\sqrt{\beta P_t}\mathbf{A}$, $\mathbf{h}_p=\textit{vec}(\mathbf{H}_{p})$, and $\mathbf{n}_p=\rho\textit{vec}
(\mathbf{G}
\mathbf{V}
\mathbf{N}_{i})
+\textit{vec}(\mathbf{N})$.
Since the elements in $\mathbf{y}_p$, $\mathbf{h}_p$, and $\mathbf{n}_p$ are all complex, the above equation can be divided into two parts, as follows:
\begin{align}\label{Re}
\mathbf{y}_p^u
=\mathbf{A}_p\mathbf{h}_p^u+\mathbf{n}_p^u,
\end{align}
and
\begin{align}\label{Im}
\mathbf{y}_p^v
=\mathbf{A}_p\mathbf{h}_p^v+\mathbf{n}_p^v,
\end{align}
where
\begin{align}
\mathbf{y}_p^u
&=\mathfrak{Re}(\mathbf{y}_{p}),~
\mathbf{y}_{p}^{v}
=\mathfrak{Im}(\mathbf{y}_{p}),\nonumber\\
\mathbf{h}^{u}_{p}
&=\mathfrak{Re}(\mathbf{h}_{p}),~
\mathbf{h}^{v}_{p}
=\mathfrak{Im}(\mathbf{h}_{p}),
\nonumber\\
\mathbf{n}_{p}^u&=\mathfrak{Re}(\mathbf{n}_{p}),~
\mathbf{n}_{p}^v=\mathfrak{Im}(\mathbf{n}_{p}).
\end{align}

Thus, the CRLB of 
$\hat{\mathbf{h}}_{p}$ can also be divided into two parts, which are shown as
\begin{align}\label{gamma-p}
\gamma_{p} =\gamma_{p}^{u}+\gamma_{p}^{v} 
=\mathbb{E}\left\{\left\|
\widehat{\mathbf{h}}^{u}_{p}
-\mathbf{h}^{u}_{p}\right\|^2\right\}
+\mathbb{E}\left\{\left\|
\widehat{\mathbf{h}}^{v}_{p}
-\mathbf{h}^{v}_{p}\right\|^2\right\}.
\end{align}
It is worth noting that since $\mathbf{n}_{p}$ is the sum of two independent AWGNs, $\mathbf{n}_{p}$ is still AWGN.
Specifically, the $\mathbf{n}_{p}$ follows the distribution of Gaussian distribution with \textbf{0} mean and $\sigma_n^2$ variance, 
where
\begin{align}
\sigma_n^2=\rho^2\sigma_i^2
\|
(\mathbf{I}_{Q_1}\otimes
\mathbf{G}
\mathbf{V})\|^2_F
+\sigma^2.
\end{align}
We first analyze the real component of Eq. (\ref{gamma-p}).
The conditional probability density function of $\mathbf{y}_{p}^{u}$ with the given $\mathbf{h}^{u}_{p}$ is
\begin{align}
p_{\mathbf{y}^{u}_{p}|\mathbf{h}^{u}_{p}}
\left(\mathbf{y}^{u}_{p};
\mathbf{h}^{u}_{p}\right)=
\frac{1}{\left(2\pi
\sigma_n^2\right)
^{MQ_1/2}}
\exp\left\{-\frac{1}{2
\sigma_n^2}\left\|
\mathbf{y}^{u}_{p}-\mathbf{A}_{p}
\mathbf{h}^{u}_{p}\right\|^2\right\}.
\end{align}
The Fisher information matrix of (\ref{Re}) can then be derived as
\begin{align}
[\mathbf{J}]_{i,j}
&=-\mathbb{E}
\left\{\frac{p_{\mathbf{y}^{u}_{p}|\mathbf{h}^{u}_{p}}
\left(\mathbf{y}^{u}_{p};
\mathbf{h}^{u}_{p}\right)}
{\partial h^{u}_{p,i}\partial h^{u}_{p,j}}\right\}
=\frac{1}{\sigma_n^2}
\left[\mathbf{A}_{p}^H\mathbf{A}_{p}\right]_{i,j},
\end{align}
where $h^{u}_{p,i}$, $h^{u}_{p,j}$ denote the $i$-th and $j$-th entry of $\mathbf{h}^{u}_{p}$. Then, the CRLB of the real part $\gamma_{p}^{u}$ is
\begin{align}
\gamma_{p}^{u}&=\mathbb{E}\left\{\left\|
\widehat{\mathbf{h}}^{u}_{p}
-\mathbf{h}^{u}_{p}\right\|^2\right\}
\geq\text{tr}
\left\{\mathbf{J}_u^{-1}\right\}
=\sigma_n^2
\mathrm{tr}
\left\{(\mathbf{A}_{p}^H
\mathbf{A}_{p})^{-1}\right\}.
\end{align}
Since $\mathbf{A}_{p}=\rho\sqrt{\beta P_t}(\mathbf{V}
\mathbf{X}_p)^T\otimes\mathbf{I}_M$, and the properties $(\mathbf{A}\otimes\mathbf{B})
(\mathbf{C}\otimes\mathbf{D}) =
(\mathbf{A}\mathbf{C})\otimes
(\mathbf{B}\mathbf{D})$, $(\mathbf{A}\otimes\mathbf{B})^H
=\mathbf{A}^H\otimes\mathbf{B}^H$, and $(\mathbf{A}\otimes\mathbf{B})^{-1}
=\mathbf{A}^{-1}\otimes\mathbf{B}^{-1}$, then $(\mathbf{A}_{p}^H
\mathbf{A}_{p})^{-1}$ can be presented as
\begin{align}
(\mathbf{A}_{p}^H
\mathbf{A}_{p})^{-1}
=\frac{1}{\rho^2 \beta P_t} \left( \left( \mathbf{X}_p^H \mathbf{V}^H \mathbf{V} \mathbf{X}_p \right)^{-1} \otimes \mathbf{I}_M \right).
\end{align}
Since $\mathrm{tr}(\mathbf{A} \otimes \mathbf{B}) = \mathrm{tr}(\mathbf{A}) \cdot \mathrm{tr}(\mathbf{B})$, thus, $\mathrm{tr}
\left\{(\mathbf{A}_{p}^H
\mathbf{A}_{p})^{-1}\right\}$ can be calculated as
\begin{align}\label{tr-A-p}
\mathrm{tr}
\left\{(\mathbf{A}_{p}^H
\mathbf{A}_{p})^{-1}\right\}
= \frac{M}{\rho^2 \beta P_t} \cdot \mathrm{tr} \left\{ \left( \mathbf{X}_p^H \mathbf{V}^H \mathbf{V} \mathbf{X}_p \right)^{-1} \right\}.
\end{align}
Assuming $\{\tau_i\}_{i=1}^{Q_1}$ is denoted as the $Q_1$ eigenvalues of the matrix of $\mathbf{X}_p^H \mathbf{V}^H \mathbf{V} \mathbf{X}_p$, we have
\begin{align}
\mathrm{tr} \left\{ \left( \mathbf{X}_p^H \mathbf{V}^H \mathbf{V} \mathbf{X}_p \right)^{-1} \right\}=\sum_{i=1}^{Q_1} \frac{1}{\tau_i},
\end{align}
then, Eq. (\ref{tr-A-p}) can be further expressed as
\begin{equation}
\mathrm{tr} \left\{ \left( \mathbf{A}_p^H \mathbf{A}_p \right)^{-1} \right\}
= \frac{M}{\rho^2 \beta P_t} \sum_{i=1}^{Q_1} \frac{1}{\tau_i},
\end{equation}

According to \cite{Lu2023Near-Field}, we have
\begin{align}
\mathrm{tr}
\left\{(\mathbf{A}_{p}^H
\mathbf{A}_{p})^{-1}\right\}
&
\geq 
 \frac{M}{\rho^2 \beta P_t}\cdot 
Q_1\Bigg(Q_1/\sum_{i=1}^{Q_1}\tau_i\Bigg)
\nonumber\\
&=
\frac{M}{\rho^2 \beta P_t}\cdot 
\frac{Q_1^2}{\mathrm{tr} \left\{  \mathbf{X}_p^H \mathbf{V}^H \mathbf{V} \mathbf{X}_p \right\}},
\end{align}
where the equality holds when $\tau_1=\tau_2=,\ldots,=\tau_{Q_1}$, namely the columns of $\mathbf{V} \mathbf{X}_p$ are orthogonal.
In this case, $\mathrm{tr} \left\{  \mathbf{X}_p^H \mathbf{V}^H \mathbf{V} \mathbf{X}_p \right\}=N^2Q_1^6$.
Finally, the CRLB of the real part of $\hat{\mathbf{h}}_{p}$ becomes
\begin{align}
\gamma_{p}^{u}=\mathbb{E}\left\{\left\|
\widehat{\mathbf{h}}^{u}_{p}
-\mathbf{h}^{u}_{p}\right\|^2\right\}
=\sigma_n^2
\frac{M}{N^2Q_1^4\rho^2 \beta P_t}.
\end{align}

Observing equations (\ref{Re}) and (\ref{Im}), it can be seen that the real part and imaginary part have the same form, namely
\begin{align}
\gamma_{p}^{v}=\gamma_{p}^{u}
=\sigma_n^2
\frac{M}{N^2Q_1^4\rho^2 \beta P_t}.
\end{align}
Thus, the CRLB of (\ref{y-hx-n}) is as folows
\begin{align}
\gamma_{p}
=\gamma_{p}^{u}
+\gamma_{p}^{v}
=2\sigma_n^2
\frac{M}{N^2Q_1^4\rho^2 \beta P_t}.
\end{align}

\section{Proposed AI-empowered CE framework}

In this section, to begin with, the basic principle of the uplink CE neural network is presented in subsection A.
Subsequently, the detailed structure of the proposed CAEformer, including encoder, MHAM, and decoder, is introduced in subsections B, C, and D, respectively. 
Finally, we evaluate the computational complexity of the proposed scheme.

\begin{figure}[h]
\centering
\includegraphics[width=3in]
{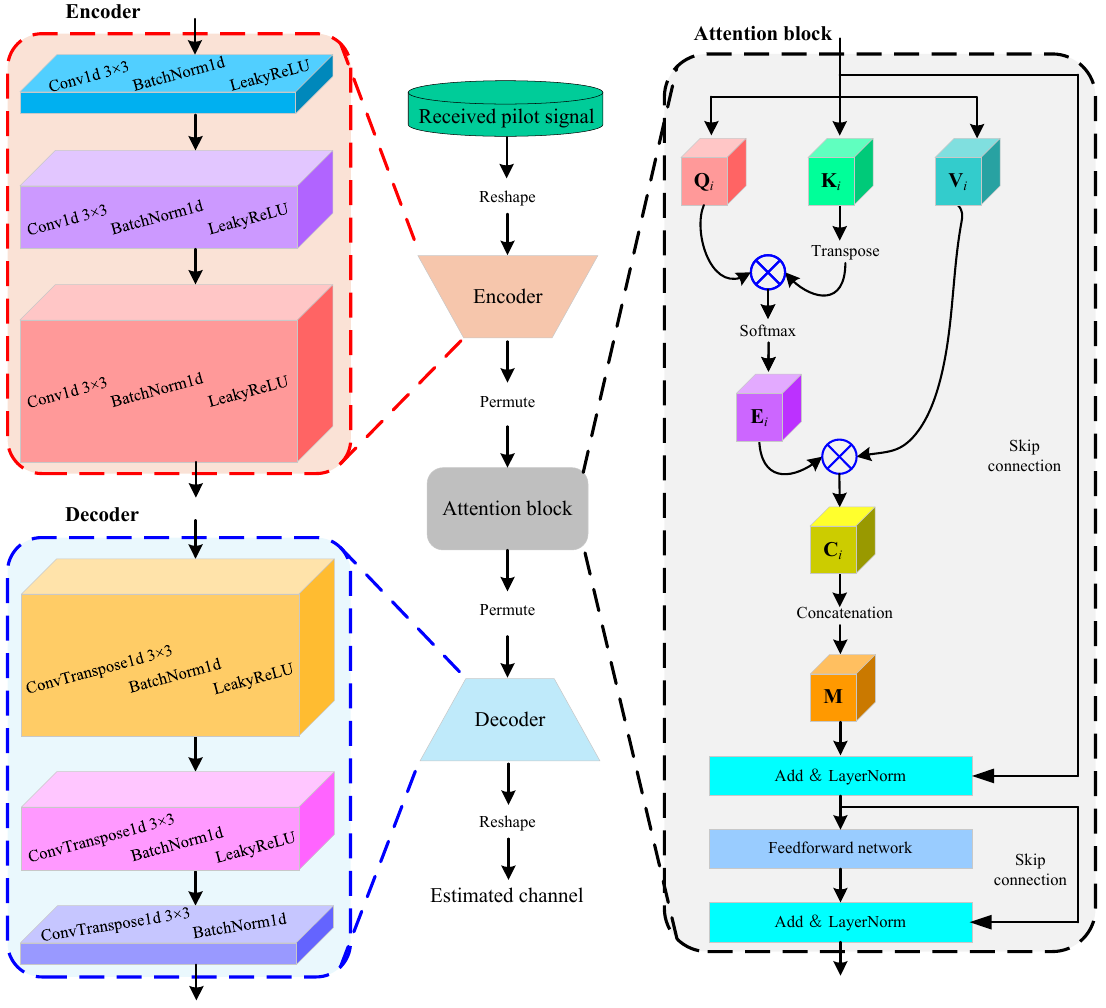}
\caption{The proposed CAEformer CE network.}
\label{transformer}
\end{figure}

\subsection{Problem Formulation}

Specifically, the proposed CAEformer network introduces a non-linear function $\boldsymbol{f}_{\boldsymbol{\omega}}$ between $\mathbf{Y}_{k}$ and $\widehat{\mathbf{H}}_{p, k,\text{CAEformer}}$, which can be mathematically expressed as
\begin{align}
\widehat{\mathbf{h}}_{p,k,
\text{CAEformer}}&
=\boldsymbol{f}_{\boldsymbol{\omega}}
(\mathbf{Y}_{k}),
\end{align}
where $\boldsymbol{\omega}$ represents weight.
In addition, $\widehat{\mathbf{h}}_{p,k,
\text{CAEformer}}
=\text{vec}(\widehat{\mathbf{H}}_{p, k,\text{CAEformer}})$ denotes the output of CAEformer.
Accordingly, the expression for the loss function is as follows
\begin{align}
\mathcal{L}(\boldsymbol{\omega})
=\frac1{D_{p,k}}\sum_{d=1}^{D_{p,k}}\parallel
\widehat{\mathbf{h}}_{p,k,
\text{CAEformer}}
-\mathbf{h}_{p,k,\mathrm{LS}}
\parallel_2^2,
\end{align}
where $D_{p,k}$ stands for the size of training dataset. The traditional LS CE scheme can be used to obtain the label $\mathbf{h}_{p,k,\mathrm{LS}}
=\mathrm{vec}(\widehat{\mathbf{H}}_{p, k})$. The objective of exploiting the proposed CAEformer network for CE is to minimize $\mathcal{L}(\boldsymbol{\omega})$ by optimizing $\boldsymbol{\omega}$, which is 
\begin{align}
\min_{\boldsymbol{\omega}}\mathcal{L}
(\boldsymbol{\omega})
&=\frac{1}{D_{p,k}}\sum_{d=1}^{D_{p,k}}\parallel
\boldsymbol{f}_{\boldsymbol{\omega}}
(\mathbf{Y}_{k})
-\mathbf{h}_{p,k,\mathrm{LS}}\parallel_{2}^{2}.
\end{align}
In each iteration $j$, $\boldsymbol{\omega}$ are updated through the following way, namely,
\begin{align}
\boldsymbol{\omega}_{j+1}
=\boldsymbol{\omega}_j
-\mu_j\mathbf{g}
(\boldsymbol{\omega}_j),
\end{align}
where $\mathbf{g}(\boldsymbol{\omega}_j)$ is the gradient vector (GV) for $\boldsymbol{\omega}_j$, and $\mu_j$ is the learning rate (LR).



\subsection{Encoder}

To effectively extract local features and contextual information from the input signal, a convolutional encoder consisting of multi-layer  one-dimensional convolution is employed to encode the original input. 
Specifically, the input signal first passes through three layers of one-dimensional convolution. Batch normalization (BN) and a non-linear activation function called Leaky Rectified Linear Unit (LeakyReLU) are connected after each convolutional layer to enhance the model's expressive capability and alleviate the gradient vanishing problem. The number of channels in each convolutional layer increases sequentially, thereby achieving layer-by-layer abstraction and compression of the input signal and extracting high-dimensional discriminative features. This module not only can effectively model the local correlations of the input but also serves as an information bottleneck for the entire network, providing a compact yet rich feature representation for the subsequent attention mechanism \cite{Li2022Lightweight}.

\subsection{Multi-Head Attention Mechanism}

To thoroughly explore long-range dependencies within sequences and enhance the capability of global context perception, a set of feature enhancement modules based on the MHAM is introduced between the encoder and the decoder.
This module first performs a transposition operation on the feature maps extracted by the convolutional encoder to meet the requirements of sequence modeling. Subsequently, it utilizes a multi-head attention layer to compute the similarities between positions in the sequence, thereby achieving global information interaction.
In addition, to improve the stability of the training process and strengthen the non-linear modeling capability, the attention module incorporates residual connections, layer normalization (LayerNorm) \cite{Singh2024Channel}, and a feedforward network. This lightweight attention enhancement unit not only effectively enhances the expressive power of features but also helps capture long-range dependencies, thus improving the model's performance in CE tasks.


For the input $\mathbf{\Upsilon}$ of the attention block, we have 
\begin{align}\label{query}
\mathbf{Q}_i=\mathbf{\Upsilon}\mathbf{\Omega}_i^q,
\end{align}
\begin{align}\label{key}
\mathbf{K}_i=\mathbf{\Upsilon}\mathbf{\Omega}_i^k,
\end{align}
\begin{align}\label{value}
\mathbf{V}_i&=\mathbf{\Upsilon}\mathbf{\Omega}_i^v,
\end{align} 
where $i=1,2,\ldots,h$ represents the $i$-th attention head, and 
$\mathbf{Q}_i$, $\mathbf{K}_i$, $\mathbf{V}_i$ stand for the trainable weight matrix for query, key, value of the $i$-th attention head. 
The corresponding transformation matrices are denoted as $\mathbf{\Omega}_i^q$, $\mathbf{\Omega}_i^k$, and $\mathbf{\Omega}_i^v$, respectively.




The attention matrix $\mathbf{E}_i$ are computed as the softmax-normalized scaled dot product of $\mathbf{K}_i^T$ and $\mathbf{Q}_i$, namely
\begin{align}\label{Softmax}
\mathbf{E}_i=\textit{Softmax}\bigg(
\frac{\mathbf{Q}_i\mathbf{K}_i^T}{\sqrt{N_k}}\bigg),
\end{align}
where $N_k$ denotes the feature dimension of the key matrix. 


The output of the attention mechanism corresponding to $i$-th head can be denoted as follows
\begin{align}\label{attention}
\mathbf{C}_i=\mathbf{E}_i\mathbf{V}_i.
\end{align}

The main advantage of multi-head attention is its ability to collectively process information from various representation subspaces. The outputs of all attention heads are concatenated to form the output of the multi-head attention layer, i.e.,
\begin{align}\label{concat-multi-head}
\mathbf{M}=\textit{concat}(
\mathbf{C}_1,\mathbf{C}_2,\ldots, \mathbf{C}_h)
\mathbf{L},
\end{align}
where $\textit{concat}(\cdot)$ denotes concatenation, $\mathbf{L}$ is a learnable weight matrix.
In different heads, the attention module focuses on various channel features \cite{Kim2023Transformer-Based}. 


\subsection{Decoder}

After feature enhancement, a convolutional decoder with a symmetric structure is designed to achieve feature restoration and reconstruction. This module consists of three layers of one-dimensional transposed convolution. BN and LeakyReLU are connected sequentially after each transposed convolutional layer to gradually upscale the feature maps and restore them to the same dimension as the input. The design objective of the decoder is to reconstruct the high-dimensional representation fused with global semantic information into channel estimation results that approximate the original input. Within the entire network architecture, the convolutional decoder plays a crucial role in mapping abstract features back to the signal space, and its reconstruction performance directly determines the estimation accuracy of the model \cite{Li2019Time-Varying}.

\begin{algorithm}
\renewcommand{\algorithmicrequire}{\textbf{Training:}}	
\renewcommand{\algorithmicensure}{\textbf{Testing:}}
\caption{Proposed CAEformer CE algorithm}
\label{EDT-algorithm}
\textbf{Initialization:} Initialize trainable parameters; raw training data $\mathbf{Y}_{k}$, and corresponding labels $\mathbf{h}_{p,k,\mathrm{LS}}$.
\begin{algorithmic}[1]
\REQUIRE Input training dataset $\mathbf{Y}_{k}$
\STATE \textbf{Encoder: } 
\FOR {each encoder layer $l \in \{1,2,\ldots,L_e\}$}
    \STATE Apply 1D convolution: $\mathbf{F}_l = \mathrm{Conv1D}_l(\mathbf{F}_{l-1})$
    \STATE Apply batch normalization 
    \STATE Apply activation function
\ENDFOR
\STATE Flatten and transpose feature map for attention module

\STATE \textbf{Attention block:}
\FOR {each head $i \in \{1,2,\ldots,h\}$}
    \STATE Compute query matrix by (\ref{query})
    \STATE Compute key matrix by (\ref{key})
    \STATE Compute value matrix by (\ref{value})
    \STATE Compute attention weights by (\ref{Softmax})
    \STATE Compute output by (\ref{attention})
\ENDFOR
\STATE Concatenate multi-head outputs by (\ref{concat-multi-head})
\STATE Apply linear projection and residual connection with LayerNorm

\STATE \textbf{Decoder:}
\FOR {each decoder layer $l \in \{1,2,\ldots,L_d\}$}
    \STATE Apply transposed convolution: $\mathbf{R}_l = \mathrm{ConvTranspose1D}_l(\mathbf{R}_{l-1})$
    \STATE Apply batch normalization
    \STATE Apply activation function
\ENDFOR
\STATE Output the well-trained CAEformer $\boldsymbol{f}_{\boldsymbol{\omega}}(\cdot)$ with well-trained parameter $\boldsymbol{\omega}$

\ENSURE
\STATE Input testing data 
\STATE Perform CE using the well-trained CAEformer network
\STATE Output estimated channel $\widehat{\mathbf{h}}_{p,k,
\text{CAEformer}}$
\end{algorithmic}
\end{algorithm}

%
%
%
%
%
%
%
%
%
%
%

\section{Simulation Results and discussions}

In this section, extensive simulations are carried out to assess the CE performance of the proposed scheme. 
In the simulation, there are a total of 9 users, with each user collecting 30000 samples, resulting in a total of 270000 samples collected. 90\% of these samples are used as the training set, while the remaining 10\% are used as the testing set.
The training process encompasses a total of 150 epochs.
The LR of the proposed CAEformer network is initialized to 1e-3, and then reduced to the half of the original level every 15 epochs.

\begin{figure}[h]
\centering
\includegraphics [width=0.41\textwidth]
{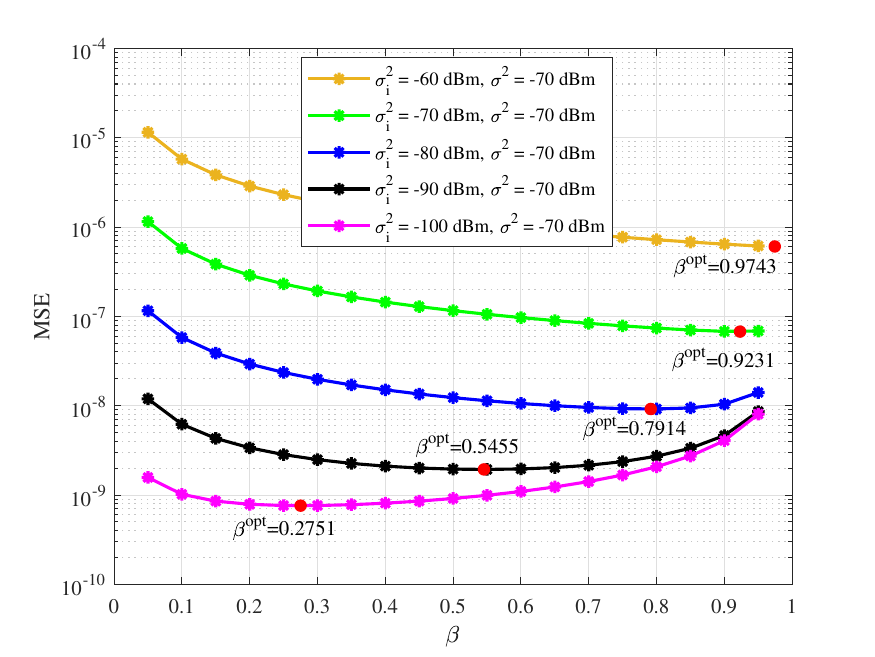}
\caption{MSE versus the PAF $\beta$ with different noise power at active IRS $\sigma_i^2$ when the noise power at BS $\sigma^2=-70$ dBm.}
\label{x_beta_variable_noise}
\end{figure}

Fig. \ref{x_beta_variable_noise} depicts the MSE versus the PAF $\beta$ with different noise power at active IRS $\sigma_i^2$ when the noise power at BS $\sigma^2=-70$ dBm.
From the figure, it can be seen that the existence of an optimal PAF achieves global minimization of MSE.
In an active IRS-assisted uplink IoT communication network, more power needs to be allocated to the IoT devices when $\sigma_i^2\geq\sigma^2$.
When $\sigma_i^2$ is slightly smaller than $\sigma^2$ by two orders of magnitude, it is still preferred to allocate power to the IoT devices.
However, when $\sigma_i^2$ is more than 3 orders of magnitude smaller than $\sigma^2$, more power should be allocated to active IRS to minimize MSE.
In addition, the specific numerical points of the optimal PAF are depicted in Fig. \ref{x_beta_variable_noise}, which is consistent with the derivation of Eq. (\ref{optimalbeta}).

\begin{figure}[h]
\centering
\includegraphics [width=0.41\textwidth]
{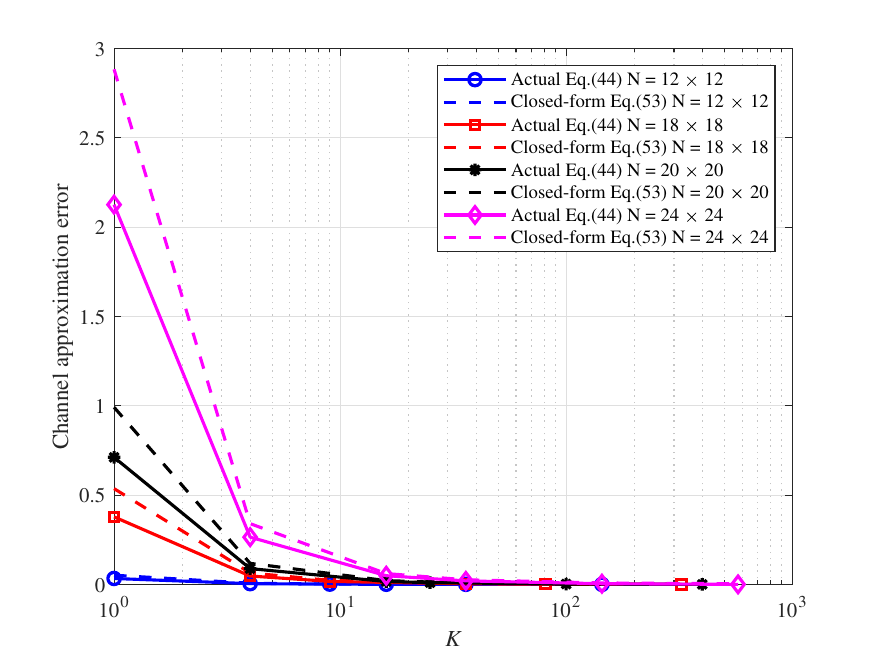}
\caption{Actual channel approximation error Eq. (\ref{appro-actual}) and closed-form channel approximation error Eq. (\ref{appro-close-form}) versus the number of sub-block $K$.}
\label{x_K_variable_N}
\end{figure}

Fig. \ref{x_K_variable_N} shows the relationship between channel approximation error  and the number of sub blocks $K$, where the solid line represents the actual channel approximation error i.e. Eq. (\ref{appro-actual}), and the dashed line represents the closed-form expression of the derived approximate channel approximation error, i.e. Eq. (\ref{appro-close-form}).
As illustrated in this figure, the channel approximation error decreases monotonically with an increasing number of sub-blocks $K$. This trend indicates that the higher the number of sub-blocks, the higher the fitting accuracy of the proposed block-based FF channel model to the actual NF channel, and this conclusion is consistent with the observation in Fig. \ref{3D-channel-phase}.
In addition, the numerical approximation between Eq. (\ref{appro-close-form}) and Eq. (\ref{appro-actual}) is significantly improved when $K$ increases, which verifies the correctness of the derivation of Eq. (\ref{appro-close-form}) and can provide a theoretical basis for the subsequent solution of the optimal number of sub-blocks $K^{\text{opt}}$.

\begin{figure}[h]
\centering
\includegraphics [width=0.41\textwidth]
{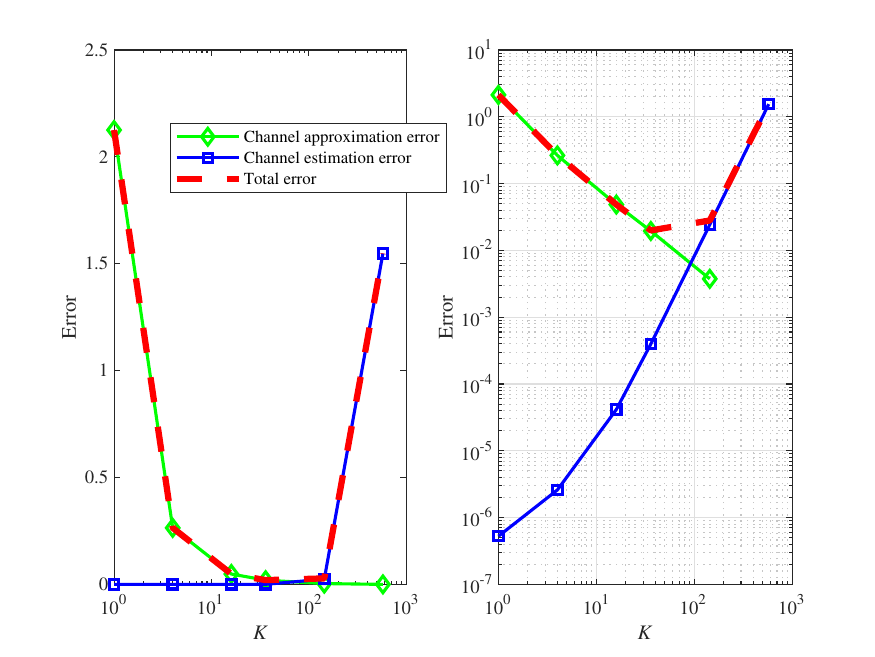}
\caption{The relationship between channel approximation error, channel estimation error, and total error versus the number of sub blocks $K$ when $N=24\times 24$.}
\label{x_K_3error}
\end{figure}

Fig. \ref{x_K_3error} depicts the relationship between channel approximation error, CEE, and total error versus the number of sub blocks $K$ when $N=24\times 24$.
From the figure, it can be seen that the CEE increases with the increase of $K$, which is consistent with the derivation of Eq. (\ref{estimation-K}).
The reason for this phenomenon is that the total pilot length of the system is limited. If the number of sub blocks increases while the pilot length remains constant, the number of pilot symbols available for each sub block will decrease, resulting in a decrease in the estimation accuracy of each sub block channel and an overall increase in estimation error.

\begin{figure}[h]
\centering
\includegraphics [width=0.41\textwidth]
{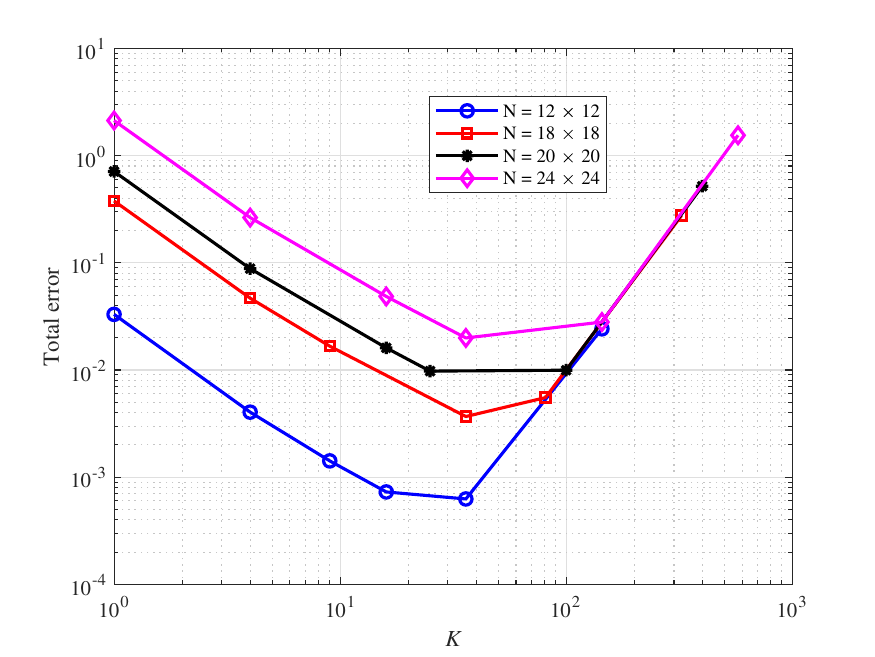}
\caption{Total error versus the number of sub-block $K$.}
\label{total_error_K}
\end{figure}

Fig. \ref{total_error_K} depicts the total error versus the number of sub-block $K$.
As shown in the figure, there exists an optimal number of sub blocks $K$ that can minimize the total error.
Although the optimal number of sub blocks $K$ may remain consistent when $N$ takes different values, there are differences in the number of elements contained within each sub block. For example, when $N=144$, the optimal number of sub blocks is $K=36$, and each sub block contains 4 elements; When $N=576$, the optimal number of sub blocks is also $K=36$, but the number of elements in each sub block increases to 16.

\begin{figure}[h]
\centering
\includegraphics [width=0.41\textwidth]
{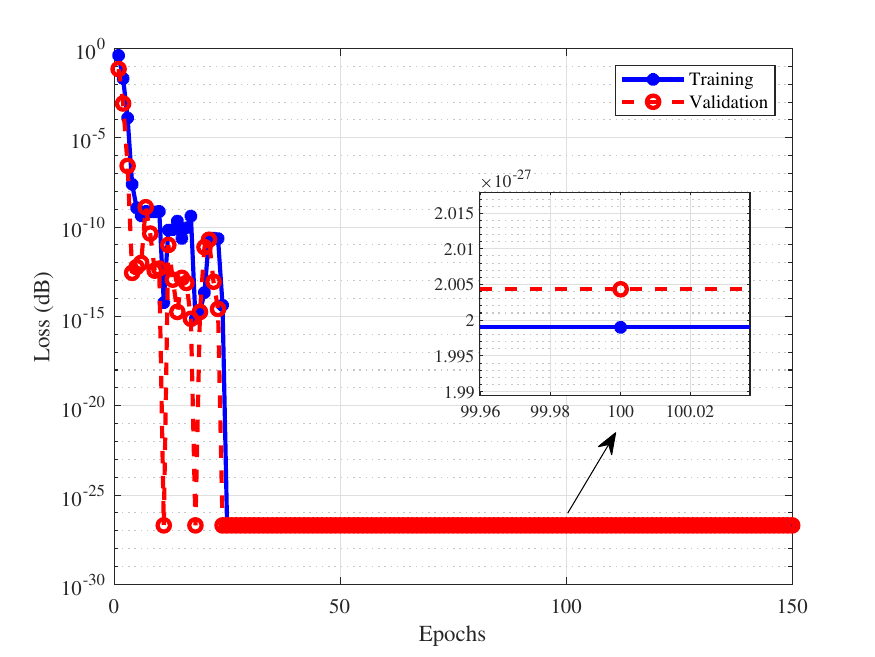}
\caption{Training and validation loss of the proposed CAEformer CE network versus the training epochs.}
\label{epoch_loss}
\end{figure}

The training and validation performance of the proposed CAEformer CE network are depicted in Fig. \ref{epoch_loss}.
As clearly demonstrated in Fig.~\ref{epoch_loss}, the performance of the designed CAEformer network has reached a stable and converged state regarding loss on both the training and validation sets after 150 epochs, with the losses remaining within acceptable bounds.

\begin{figure}[h]
\centering
\includegraphics [width=0.41\textwidth]
{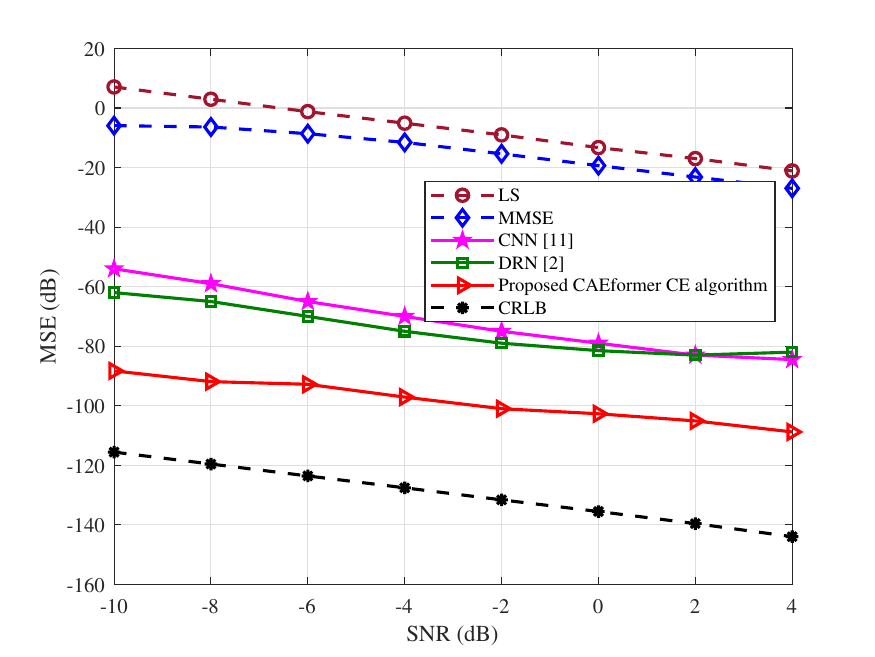}
\caption{The MSE performance versus the SNR.}
\label{SNR_MSE}
\end{figure}

The feasibility and effectiveness of the proposed CAEformer algorithm are validated by comparing its MSE performance with those of other existing methods across different SNRs, as illustrated in Fig. \ref{SNR_MSE}.
Evidently, owing to the non-linear mapping capability provided by the MHAM, the proposed CAEformer network is capable of learning the unique channel characteristics of the received pilot signals, thereby achieving higher CE accuracy compared to the CNN and DRN schemes proposed in \cite{Wang2024Power} and \cite{Wang2024Enhanced}.

\section{Conclusion}

In this paper, CE for IRS-aided uplink hybrid-field communications has been investigated.
Firstly, an IRS with $N$ elements was partitioned into $K$ small sub-blocks, such that the NF channel between the user and the IRS could be approximated by a FF channel model between the user and each sub-block. 
This approach significantly reduced the complexity of channel modeling while decreasing the parameter estimation dimension from $N$ to $K$. Additionally, the relationships among channel approximation error, CE error, and $K$ were mathematically derived, enabling the determination of the optimal $K$ that minimized the total error.
Secondly, to address the power consumption challenges of active IRS, the optimal PAF for minimizing the total error was derived. The theoretical performance limit of the proposed algorithm was validated using the CRLB.
Finally, to further reduce pilot overhead, a lightweight CE framework named CAEformer was proposed, which integrated convolutional feature extraction with a MHAM. 
By synergistically modeling both local and global channel characteristics while maintaining low computational complexity, the propsoed CAEformer achieved a favorable balance between estimation accuracy and efficiency. Simulation results demonstrated that the propsed CAEformer significantly outperformed conventional methods such as LS, MMSE, CNN \cite{Wang2024Power}, and DRN \cite{Wang2024Enhanced} in terms of CE accuracy, while effectively balancing computational complexity and estimation performance.



\end{document}